\documentclass[pdflatex,sn-mathphys-num,iicol]{sn-jnl}

\usepackage{graphicx}%
\usepackage{multirow}%
\usepackage{amsmath,amssymb,amsfonts}%
\usepackage{amsthm}%
\usepackage{mathrsfs}%
\usepackage[title]{appendix}%
\usepackage{xcolor}%
\usepackage{textcomp}%
\usepackage{manyfoot}%
\usepackage{booktabs}%
\usepackage{algorithm}%
\usepackage{algorithmicx}%
\usepackage{algpseudocode}%
\usepackage{listings}%
\usepackage[caption=false]{subfig}
\usepackage{siunitx}
\usepackage{braket}
\usepackage[final]{changes}

\theoremstyle{thmstyleone}%
\theoremstyle{thmstyletwo}%

\theoremstyle{thmstylethree}%

\newcommand{\PrYSO}{Pr$^\textup{3+}$:Y$_2$SiO$_5$}

\begin{document}
\title{Distribution of light-matter quantum correlations with a temporally multiplexed solid-state quantum memory array}

\author[1]{\fnm{Aya} \sur{Mneimneh}}
\equalcont{These authors contributed equally to this work.}
\author[1]{\fnm{Susana} \sur{Plascencia}}
\equalcont{These authors contributed equally to this work.}
\author[1]{\fnm{Manuel} \sur{Gundin}}
\author[1]{\fnm{Jonathan} \sur{Hänni}}
\author[1]{\fnm{Samuele} \sur{Grandi$^\ddagger$}
}
\author*[1]{\fnm{Markus} \sur{Teller}}\email{markus.teller@icfo.eu}
\author[1,2]{\fnm{Hugues} \sur{de Riedmatten}}

\affil[1]{\orgdiv{ICFO-Institut de Ciencies Fotoniques}, \orgname{The Barcelona Institute of Science and Technology}, \orgaddress{\city{Castelldefels (Barcelona)}, \postcode{08860}, \country{Spain}}}

\affil[2]{\orgdiv{ICREA}, \orgname{Institucio Catalana de Recerca i Estudis Avançats}, \orgaddress{\city{Barcelona}, \postcode{08015}, \country{Spain}}}
\date{\today}


\abstract{Multiplexed quantum memories increase the entanglement distribution rate in long-distance quantum repeater architectures by harnessing storage in several degrees of freedom.
Here, we report on the distribution of light-matter quantum correlations using an array of time-multiplexed solid-state quantum memories.
We store telecom-heralded single photons sequentially in up to ten memory cells using the full atomic frequency comb protocol with on-demand read-out in a {\PrYSO} crystal.
Leveraging both spatial and temporal multiplexing, we demonstrate quantum correlations between the telecom photon and up to 60 spatio-temporal modes of the quantum memory array. 
We then transmit the heralding telecom photon over \SI{39.1}{\kilo\meter} of deployed optical fiber in the Metropolitan Area of Barcelona.
In a realistic scenario where the generation rate is limited by the two-way communication time, we show that up to $\SI{15}{\%}$ of the \SI{393}{\micro\second} round-trip communication time is filled with communication trials, leading to a 60-fold enhancement in the rate of detected telecom photons correlated with the quantum memory array, compared to a single-mode memory.
With increased storage times and efficiencies, our multiplexed quantum memory array will constitute the backbone of a long-distance quantum network, establishing remote entanglement at high rates.
}

\maketitle

\footnotetext[3]{Present address: Arq Quantum Technologies, 08042 Barcelona.}

Quantum repeaters have been proposed to distribute entanglement among distant parties~\cite{Briegel1998, Kimble2008, Sangouard2011, Wehner2018}, overcoming losses in quantum channels to constitute the backbone of large-scale quantum networks.
A practical quantum internet implementation~\cite{Kimble2008, Wehner2018} requires, however, not only establishing entanglement over long distances, but also its distribution at high rates.
For near-term quantum repeater architectures relying on heralded entanglement between remote quantum nodes \cite{Sangouard2011}, the communication time imposes a fundamental limit on entanglement rates: subsequent entanglement generation trials are delayed by the photon propagation time and by the return of the classical signal heralding a successful attempt~\cite{Simon2007}.
Multiplexed quantum repeaters overcome this limitation by attempting to establish entanglement in independent modes during the communication time, thereby ensuring increasingly continuous operation of the repeater and linearly enhancing the entanglement distribution rate with the number of modes addressed during this time~\cite{Ortu2022b}.

Atomic ensembles, both in atomic clouds and solid-state matrices, combine high multiplexing capacity with on-demand storage and retrieval of quantum correlations, essential requirements for quantum repeater nodes~\cite{Simon2007}.
Large numbers of spatial modes have been reported in cold atomic ensembles~\cite{Lan2009, Pu2017, Tian2017, Chang2019}, with recent experiments demonstrating enhanced distribution rates of quantum correlations through $\SI{12}{\kilo\meter}$ of in-laboratory fiber~\cite{Zhang2024a}.
Time multiplexing has also been achieved in a cold atomic quantum memory \cite{Heller2020}. 
However, combining spatial multiplexing with other degrees of freedom is challenging in cold atomic ensembles, which limits the total number of achievable modes.
In contrast, solids doped with rare-earth ions offer intrinsic temporal multiplexing e.g. by using the atomic-frequency comb protocol (AFC)~\cite{Afzelius2009, Lago-Rivera2021, Businger2022, Zhu2026}, with recent demonstrations combining this approach with spatial multiplexing to achieve a large number of modes~\cite{Gundogan2012, Yang2018, Teller2025a, Ou2025}, and with frequency multiplexing \cite{Sinclair2014, Seri2019, Tateishi2026}.
Despite these advantages, a solid-state quantum node featuring non-classical correlations and spatial multiplexing has not yet been demonstrated.

In this work, we close this gap by combining a solid-state quantum memory array (QMA) and an entangled photon-pair source to form a spatially and temporally-multiplexed quantum node.
Quantum correlations are generated by a non-degenerate cavity-enhanced spontaneous parametric down-conversion (cSPDC) source, emitting pairs of entangled signal and telecom idler photons~\cite{Simon2007, Fekete2013,Rielander2016}.
The QMA~\cite{Teller2025,Teller2025a} features ten cells of AFC spin-wave memories~\cite{Afzelius2009}, each with six temporal modes and on-demand read-out.
We report telecom-heralded single-photon storage in the QMA, demonstrating light-matter quantum correlations in a total of $60$ spatio-temporal modes with cross-correlations exceeding the classical limit for each memory cell.

A key challenge is the deployment over a metropolitan network~\cite{Rakonjac2023, Stolk2024, Knaut2024, Liu2024, Zhu2026}.
Here, we harness the telecom compatibility of our quantum node to distribute quantum correlations across a metropolitan quantum network testbed, propagating idler photons through $\SI{39.1}{\kilo\meter}$ of deployed fiber to a remote detector station, corresponding to a round-trip communication time of $\SI{393}{\micro\second}$.
We operate in a realistic scenario where the generation rate of light-matter correlations is limited by the round-trip communication time.
Due to the multiplexing capabilities of our node, we show that up to a total of $\SI{15}{\%}$ of the communication time is filled with useful heralding events, a ten-fold increase with respect to a single AFC memory and a $60$-fold increase over a system without multiplexing.
This increase leads to a direct enhancement of the correlated telecom photon detection rate and coincidence rate at the remote station.

\section{Results}
\subsection{Experimental setup}

\begin{figure*}[]
    \centering
    \includegraphics[width=1\textwidth]{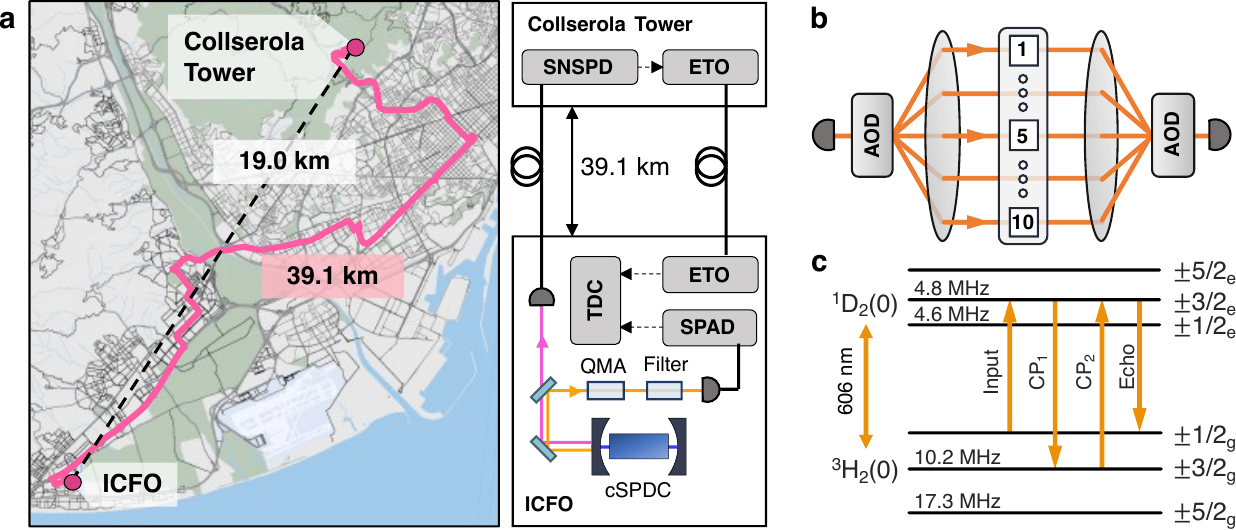}
    \caption{\textbf{Experimental setup}. a) Quantum network testbed.
    Signal photons generated by the cSPDC source in ICFO are stored in the QMA and, upon on-demand retrieval, are spectrally filtered, detected by a single-photon avalanche detector (SPAD) and recorded by a time-to-digital converter (TDC).
    Telecom idler photons propagate to the Collserola Tower through the deployed fiber, where they are detected by superconducting nanowire single-photon detectors (SNSPDS). 
    A pair of optical transducers (ETOs) sends the timestamps of the idler detections back to ICFO.
    b) The solid-state quantum memory array is realized with acousto-optical deflectors (AODs), which multiplex and demultiplex signal photons to 10 individual quantum memory cells of the {\PrYSO} crystal.  
    c) Electronic level structure of \PrYSO. 
    The AFCs are prepared on the $\SI{606}{\nano\meter}$ transition from $\ket{g}=\pm1/2_g$ to $\ket{e}=\pm3/2_e$, corresponding to the signal input and echo of the quantum memories. Control pulses (CPs) implement on-demand storage and read-out by transferring population between $\ket{s}=\pm3/2_g$ and $\ket{e}=\pm3/2_e$.
    }
    \label{fig:1}
\end{figure*}

Our quantum network testbed consists of a multiplexed quantum node at ICFO, a detector station at the Collserola Tower, located just outside of Barcelona, and a deployed fiber connecting the two locations, as shown in Fig.~\ref{fig:1}a. 
The quantum node comprises the QMA and a cSPDC source based on a periodically poled potassium titanyl phosphate (ppKTP) crystal integrated into a bow-tie optical cavity\cite{Fekete2013}. 
Pumping of the crystal with light at $\SI{436}{\nano\meter}$ generates signal photons at $\SI{606}{\nano\meter}$ and idler photons in the telecommunication C-band at $\SI{1552}{\nano\meter}$.
Resonance of the cavity with both signal and idler enhances the photon-pair emission and ensures the signal photon is generated in a $\SI{1.40(2)}{\mega\hertz}$ cavity mode, compatible with the $\SI{4}{\mega\hertz}$ quantum memory bandwidth.
We pump the cSPDC with a $\SI{500}{\micro\watt}$ continuous-wave laser, leading to a detected idler rate at ICFO of $\SI{6000}{\second^{-1}}$ and a $40\%$ heralding efficiency in fiber.
Heralded signal photons emitted by the cSPDC source are sequentially stored in the QMA, leading to a time-to-spatial conversion.
Upon on-demand retrieval, they are spectrally filtered and detected by a single-photon avalanche photodiode (SPAD)  (see Methods).
A time-to-digital converter (TDC) records the timestamps of these detection events. 

Idler photons travel to the remote detector station through the deployed fiber link.
Hence, each memory cell is correlated to temporal modes at different points in the fiber.
We detect a telecom idler rate of $R_\mathrm{T}=\SI{300}{\second^{-1}}$ due to the $\SI{10.01}{\dB}$ losses along the $L=\SI{39.1}{\kilo\meter}$ fibre length, and additional losses from fiber mating connections in ICFO and the Collserola Tower, which yield a total transmission of about $5\%$.
In the remote station, idler photons are spectrally filtered to remove noise coupled to the fiber and detected by superconducting nanowire single-photon detectors (SNSPDs).
Noise in this channel is limited only by detector dark counts.
A custom-built electrical-to-optical (ETO) transducer converts the detector signal into an optical pulse.
We send this pulse to ICFO through a second deployed fiber, convert it into a TTL signal and record it in the same TDC used for the signal photons, thus avoiding the need for clock synchronization between locations.
The total round-trip communication time imposed by the link is ${T_\mathrm{com}=\SI{393}{\micro\second}}$. 


The QMA is implemented in a {\PrYSO} rare-earth ion doped crystal by two acousto-optic deflectors (AODs) placed in the focal plane of two lenses in a $4f$ configuration, with the quantum memory crystal in the center as illustrated in Fig.~\ref{fig:1}b. 
Separate optical paths with additional AODs are used to simultaneously prepare the full array and to implement on-demand control.
The AFC is prepared on the $\SI{606}{\nano\meter}$ transition from ${\ket{g}=\pm1/2_g}$ to ${\ket{e}=\pm3/2_e}$, between the {\PrYSO} states $^3H_4(0)$ and $^1D_2(0)$.
The energy levels of the system are shown in Fig.~\ref{fig:1}c.
Without on-demand control, single photons stored in one of the ten AFC memory cells are reemitted after a delay $\tau$ predetermined by the structure of the frequency comb. 
Single photons can be stored in several temporal modes and retrieved in a first-in-first-out fashion due to the intrinsic temporal multimodality of the AFC protocol~\cite{Afzelius2009}.

On-demand storage is implemented with a control pulse (CP) that transfers the excited-state population $\ket{e}$ after photon absorption to the spin state $\ket{s}=\pm3/2_g$ \cite{Afzelius2010a}, thus freezing the AFC rephasing.
After a controllable storage time T$_s$, a second CP transfers the population back to the excited state, where the AFC rephasing continues and coherent reemission of the stored photons takes place after a total time T$_s+\tau$, thus implementing on-demand read-out.
A comprehensive description of the QMA is detailed in Refs.~\cite{Teller2025,Teller2025a}.

\subsection{Experimental sequence}

\begin{figure*}[]
    \subfloat{
        \label{fig:2a}
    \includegraphics[width=\textwidth]{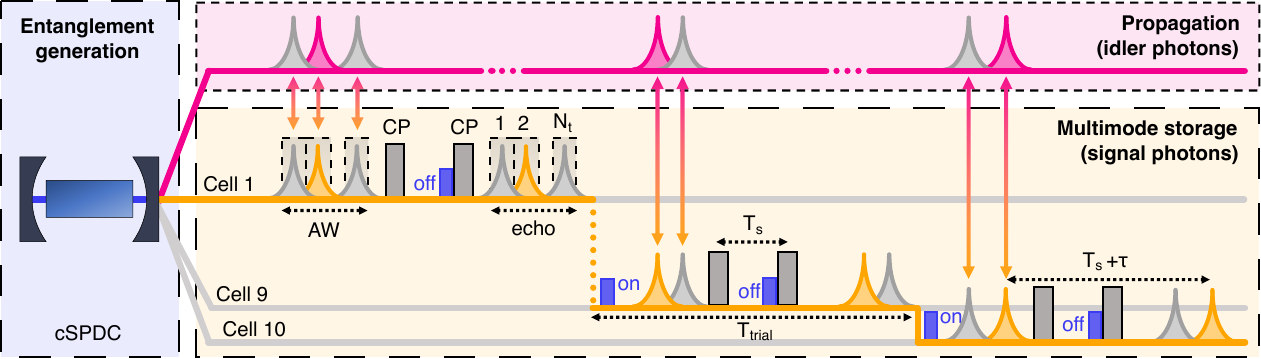}
     }

    \caption{\textbf{Experimental sequence}. The cSPDC source emits pairs of signal and idler photons within an acceptance window (AW) of \SI{6}{\micro\second} corresponding to $N_t$ temporal modes. 
    While the idler photon travels to the detector, the first cell $N_\mathrm{s}=1$ of the QMA absorbs the signal photon. Subsequently, a CP is applied, and the cSPDC pump is switched off. After a storage time $T_s=\SI{8}{\micro\second}$, a second CP is applied and leads to re-emission of the signal photon after $T_\mathrm{s}+\tau = \SI{18}{\micro\second}$. The cSPDC is switched on again, and quantum storage continues with the next memory cell. These steps are repeated until storage has been attempted with all memory cells. 
    }
    \label{fig:2}
\end{figure*}

The sequence of single-photon storage is illustrated in Fig.~\ref{fig:2} and begins by simultaneously preparing an AFC with $\tau= \SI{10}{\micro\s}$ in all memory cells~\cite{Teller2025a}. 
Quantum storage commences by addressing the first memory cell ($N_\mathrm{s}=1$). 
Continuous pumping of the cSPDC leads to probabilistic emission of photon pairs within a $\SI{6.0}{\micro\second}$ acceptance window (AW), in one of the $N_\mathrm{t}$ temporal modes.
The signal photon enters the quantum memory cell, while the corresponding telecom idler travels to a detector.
Immediately after the AW, a $\SI{3.5}{\micro \s}$ CP is applied, and the SPDC pump is turned off to avoid noise generated by the pump field to coincide later with the retrieved photon.
After a spin-wave storage time $T_\mathrm{s} = \SI{8}{\micro \s}$, a second CP leads to the emission of the stored photons during the following $\SI{6}{\micro \s}$ (echo).
The total storage time is $T_s+\tau = \SI{18}{\micro \second}$, with the switching times of the AODs increasing the total trial time required to address each memory cell to $T_\mathrm{trial}=\SI{42.7}{\micro \s}$.
After retrieval of the signal photon, the cSPDC is switched on, and quantum storage continues with the next memory cell.
This is repeated until storage has been attempted in the full QMA, resulting in a total sequence duration of ${T_\mathrm{seq}=\SI{427}{\micro \s}}$.
To be compatible with long-distance transmission, the full sequence is unconditional to the detection of an idler photon.
The sequence is repeated 400 times per cryostat duty cycle ($\SI{0.7}{\second}$), corresponding to over 5700 storage attempts per second, whereas the QMA is prepared only once per cycle.

\subsection{Heralded single-photon storage}
\begin{figure*}[]
    \subfloat{
        \includegraphics[width=1\textwidth]{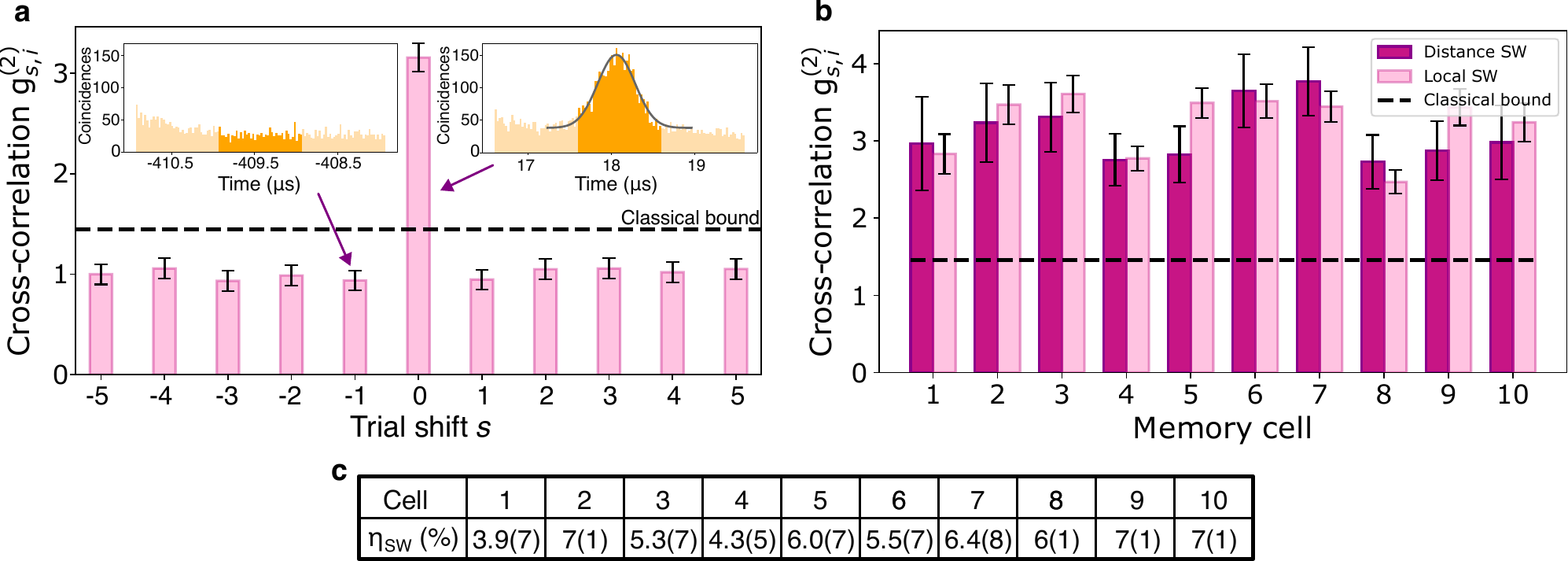}
        \label{fig:3a}
        }
    \subfloat{
        \label{fig:3b}
     }
         \subfloat{
        \label{fig:3c}
     }
    \caption{\textbf{Multimode single-photon storage.} a) Cross-correlations of idlers and signals detected in a storage trial shifted by $s$ from the idler. The insets show coincidence histograms of the signals and idlers. The peak in the right inset corresponds to the spin-wave echo of the retrieved signal photons. This echo is absent in the coincidence histogram in the left inset with $s=-1$. The highlighted area indicates the detection window $T_\mathrm{DW}$. b) Cross-correlations $g^{(2)}_{s,i}$ for spin-wave storage with idlers detected locally and at distance with $T_{DW} = \SI{1}{\micro\second}$. The dashed line indicates classical bounds $b_\mathrm{l} = 1.42(16)$ and $b_\mathrm{d} = 1.45(18)$ of the local and link measurements. c) Efficiencies of the single-photon storage per memory cell.}
    \label{fig:3}
\end{figure*}

Before proceeding to experiments with the fiber link, we characterize the single-photon storage performance of the quantum node with the idler events detected locally using the sequence described in Fig.~\ref{fig:2}.
We record timetags of signal and idler events over $\SI{6.4}{\hour}$ and correlate these detections as a function of the delay time between the heralding idler and the signal retrieved from the memory.
In Fig.~\ref{fig:3a}, right inset, we plot the coincidences as a function of the time between the two events, adding up the coincidences from all spatial modes.
We find a coincidence peak at $\SI{18}{\micro\second}$, consistent with the total storage time, with a FWHM of $\SI{531(13)}{\nano\second}$ from a Gaussian fit.  
For details on the photon linewidth, see the Supplemental Materials.
To verify that this peak represents the retrieved signal photons, we correlate the idlers with signal photons retrieved from the same quantum memory cell in a storage trial $s$ attempted $s\cdot T_\mathrm{seq}$ before or after the herald.
In Fig.~\ref{fig:3a}, left inset, we plot the coincidence histogram for $s=-1$.
The absence of a spin-wave echo confirms that for $s=0$ the peak corresponds to the retrieved signal photons. 

We assess the quality of photon storage with the temporal cross-correlation $g_{s,i}^{(2)} = p_\mathrm{s,i}/p_\mathrm{s}p_\mathrm{i}$ calculated with the probability to detect a coincidence ($p_\mathrm{s,i}$) and the probabilities to detect uncorrelated idlers and signals ($p_\mathrm{i}$ and $p_\mathrm{s}$) within a detection window $T_\mathrm{DW}$~\cite{Seri2017}.
Considering a window of $T_\mathrm{DW}=\SI{1}{\micro\second}$, highlighted in the insets of Fig.~\ref{fig:3a} and equal to about $\SI{90}{\%}$ of the signal photon, we evaluate $g_{s,i}^{(2)}$ for trial shifts $s=\{-5,5\}$ and plot the resulting cross-correlations in Fig.~\ref{fig:3a}.
See methods for details on the calculation of $p_\mathrm{i}$ and $p_\mathrm{s}$.
For $s=0$ we measure a cross-correlation of $g_{s,i}^{(2)} = 3.2(1)$.
This value violates the classical bound given by the Cauchy-Schwarz inequality $b_\mathrm{l}=\sqrt{g_{s,s}^{(2)}g_{i,i}^{(2)}}=1.4(1)$, with $g_{s,s}^{(2)}$ ($g_{i,i}^{(2)}$) the unheralded signal (idler) autocorrelation (see Supplementary Materials) by $16$ standard deviations, confirming quantum correlations between the telecom photons and the QMA. For $s\neq 0$, the cross-correlation is on average $1.0(1)$, significantly below the classical bound, as expected due to the absence of correlations between different storage trials.
Note that the uncertainties are calculated by assuming Poisson statistics of the detection events.

So far, we have treated the QMA as a single quantum memory.
We now evaluate the storage performance per memory cell, calculating $g_{s,i}^{(2)}$ for each of them individually.
We plot these values and the Cauchy-Schwarz bound $b_\mathrm{l}$ in Fig.~\ref{fig:3b} (labelled Local SW).
The cross-correlations range from $2.5(2)$ to $3.6(2)$ and violate the classical bound for all cells by at least $4.8$ standard deviations.
To assess the efficiency of the single-photon storage, we compare the coincidence probability $p_\mathrm{s,i}$ with the probability obtained from a separate measurement where the signal photons passed through a transparency window in the QMA, thus without AFC preparation or storage (see supplementary material).
The photon storage and retrieval efficiencies are listed in Fig.~\ref{fig:3c} per memory cell and range between $3.9(7)$ and $\SI{7(1)}{\percent}$. 
We attribute this variation to differences in the AFC efficiencies caused by imperfect alignment of the 4f system and drifts of the CP beam during the measurements~\cite{Teller2025a}.
Note that these efficiencies are calculated considering only about $\SI{90}{\percent}$ of the retrieved signal photon is within $T_\mathrm{DW}$, hence decreasing the result.

\subsection{Distribution of quantum correlations over the metropolitan fiber network}
We now turn to experiments with the deployed fiber link.
We repeat the previous measurements using the same experimental sequence but sending the idler photon across the deployed fiber.
Due to the communication time $T_\mathrm{com}$, the signal photons are retrieved before the idler is detected at the remote station.
We acquire photon statistics over $\SI{25}{\hour}$ with up to $\SI{827.9(3)}{min^{-1}}$ detected rate of correlated idler photons and evaluate the cross-correlation between idlers and retrieved photons per memory cell.
The resulting correlations are compared in Fig.~\ref{fig:3b} to the classical bound and to the values obtained from the previous measurements without the link.
The values obtained with the link violate the classical bound of ${b_\mathrm{d} = 1.5(1)}$ for all cells for at least $2.2$ standard deviations.
Moreover, the cross-correlations obtained with and without the link coincide within their errors, from which we conclude that noise entering the link is effectively filtered before reaching the SNSPDs. 

In the experiments presented here, the cross-correlations are limited by fluorescence noise from the CPs detected during the detection window.
Higher values of $g_{s,i}^{(2)}$ can be obtained with shorter detection windows, at the expense of coincidence rates.
This trade-off is particularly relevant for future quantum network applications, in which target fidelities are prioritized over the time required to establish long-distance entanglement to execute an application~\cite{DelleDonne2025}.
We analyze this trade-off in cross-correlations and coincidence rate as a function of $T_\mathrm{DW}$ for windows between $200$ and $\SI{1200}{\nano\second}$, equivalent to $25$ and $\SI{93}{\percent}$ of the retrieved photon respectively.
For each window, we average the cross-correlations and sum the coincidence rates of all cells.  
\begin{figure}[]
        \includegraphics[width=0.5\textwidth]{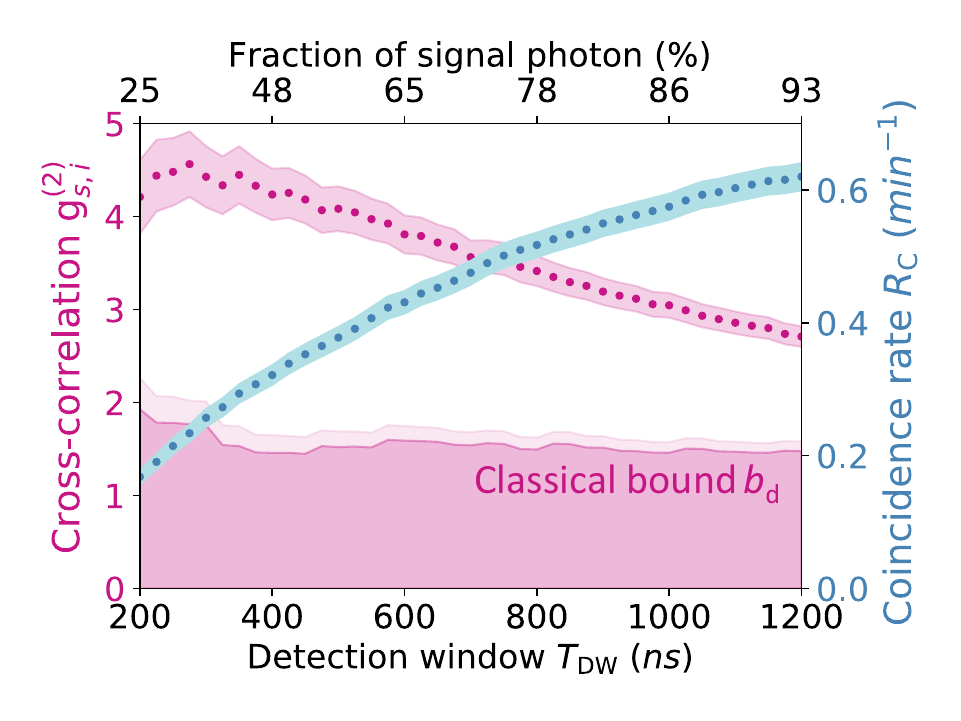}
    \caption{\textbf{Cross-correlation $g^{(2)}_{s,i}$ and coincidence rate $R_\mathrm{C}$ versus detection window $T_\mathrm{DW}$.} The averaged cross-correlation and the total coincidence rate are plotted as a function of the detection window $T_\mathrm{DW}$ with the corresponding fraction of photon within $T_\mathrm{DW}$. The classical bound $b_\mathrm{D}$ is indicated as solid line and area below. The light area above marks the uncertainty of $b_\mathrm{D}$.}\label{fig:4}

\end{figure}

In Fig.~\ref{fig:4}, we plot the average cross-correlation and the total coincidence rate $R_\mathrm{C}$.
The classical bound $b_\mathrm{d}$ is indicated as a shaded area with the light area representing the uncertainty of $b_\mathrm{d}$.
With increasing $T_\mathrm{DW}$, the detected coincidence rate $R_\mathrm{C}$ increases from $0.17(1)$ to $\SI{0.62(2)}{\min^{-1}}$ by more than a factor of three.
The average cross-correlation decreases from $4.2(4)$ to $2.7(1)$, while violating the classical bound for all values of $t_\mathrm{DW}$ for at least $2.8$ standard deviations and violations up to $10.1$.
We therefore conclude that the QMA preserves the quantum correlations between signal and idler photons for all values of $T_\mathrm{DW}$, including windows that cover more than $\SI{90}{\percent}$ of the retrieved signal. It is important to note that when reducing $T_\mathrm{DW}$, while the rate of coincidences decreases, the heralding rate remains constant since more temporal modes can be used. For example, for $T_\mathrm{DW}=\SI{300}{\ns}$ where $g^{(2)}_{s,i}$ is maximum, the total number of temporal modes would be $20$, leading to $200$ spatio-temporal modes.     

We now study the enhancement of the distribution rate due to our multiplexed quantum memory array. With ${T_\mathrm{AW}=\SI{6}{\micro\second}}$ of acceptance window per memory cell, the QMA fills $T_\mathrm{com}$ with ${T_\mathrm{tot}=N_\mathrm{s}T_\mathrm{AW}=\SI{60}{\micro\second}}$ of storage attempts.
To assess the enhancement, we divide $T_\mathrm{tot}$ in $N=60$ spatio-temporal modes of $T_\mathrm{DW}=\SI{1}{\micro\second}$ each.
Each memory cell therefore contains $N_\mathrm{t}=6$ temporal modes and fills $\SI{1.5}{\percent}$ of $T_\mathrm{com}$.
Starting with the first spatio-temporal mode $i=1$, i.e. memory cell one and temporal mode one, we plot the telecom photon detection rate $R_\mathrm{T} = \sum_{i=1}^{N} R_{\mathrm{T},i}$ and coincidence rates $R_\mathrm{C} = \sum_{i=1}^{N} R_{\mathrm{C},i}$ for $N$ increasing from  $N=1$ to $60$, equivalent to a filling of the communication time $F = NT_\mathrm{DW}/T_\mathrm{com}$ ranging from $\SI{0.25}{\percent}$ to $\SI{15}{\percent}$.

\begin{figure}[]
    \includegraphics[width=0.5\textwidth]{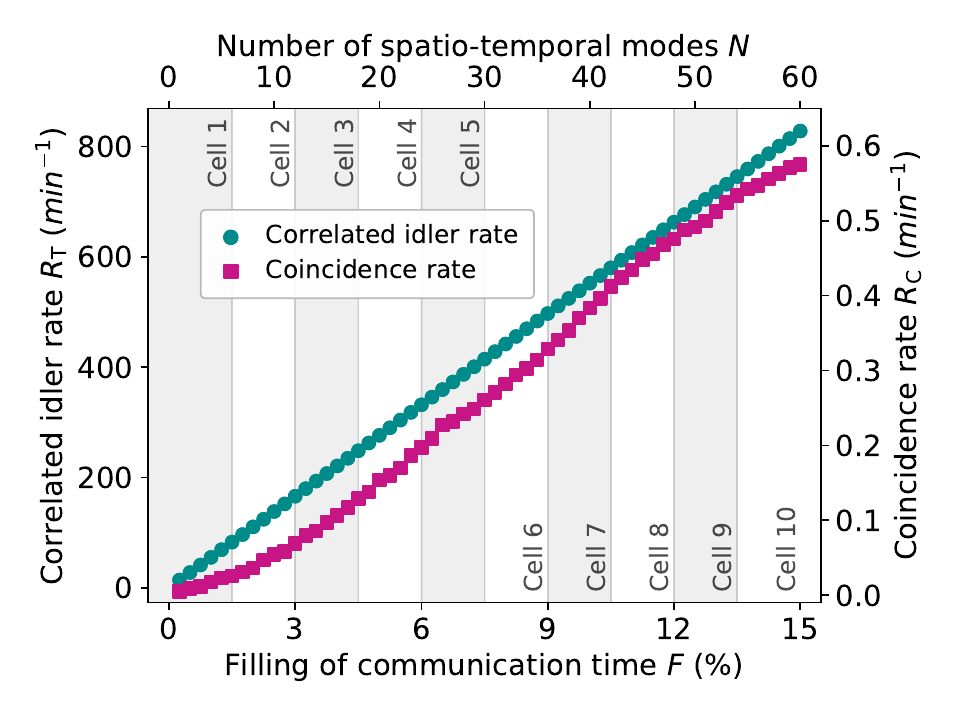}\label{fig:4b}
    \caption{\textbf{Enhancement in distribution rates of quantum correlation.} Heralding and coincidence rates as a function of the filled communication time $F$ and spatio-temporal modes $N$. 
    \label{fig:5}
    }
    
\end{figure}
The resulting rates are plotted in Fig.~\ref{fig:5}. The detection rate of correlated telecom photons  $R_\mathrm{T}$ increases linearly from $\SI{13.9(1)}{\min^{-1}}$ of a single temporal mode $N_\mathrm{t}=1$ stored in cell one to $\SI{827.9(3)}{\min^{-1}}$ for $N_\mathrm{t}=6$ temporal modes stored in each of the ten memory cells. Comparing ${R_\mathrm{T} = \SI{82.7(1)}{\min^{-1}}}$  obtained with a single memory cell storing $N_\mathrm{t}=6$ to the heralding rate obtained of the full QMA with $N=60$, we find a ten-fold enhancement due to the spatial multiplexing.
The coincidence rate increases with $N$, although not linearly due to the varying efficiencies, as listed in Fig.~\ref{fig:3a}. 
Here we calculate an enhancement of $7$ with respect to the most efficient cell storing $N_\mathrm{t}=6$ temporal modes.
In comparison to a quantum memory without multiplexing, i.e., with $N=1$ and a pulsed source of $\SI{1}{\micro \s}$-long photon-pairs, the measured detection rate of correlated telecom single photons with multiplexing is $60$ times higher.
We note that only about \SI{92}{\%} of the total sequence time ${T_\mathrm{seq}=\SI{427}{\micro \s}}$ falls within the communication time ${T_\mathrm{com}=\SI{393}{\micro \s}}$, however, all acceptance windows of telecom photons are within $T_\mathrm{com}$. 
We emphasize that the detected rate of correlated telecom photons in our experiment is comparable (a factor two lower) to the expected heralding rate of a single-click entanglement scheme between two spatially-multiplexed quantum nodes separated by ${2L=\SI{78.2}{\kilo\meter}}$~\cite{Lago-Rivera2021, Haenni2025}.
This highlights the prospects of this platform for long-distance entanglement distribution at high rates.

\section{Discussion}

In conclusion, we demonstrated quantum correlation between a telecom photon and an array of 10 temporally multiplexed solid-state quantum memories with on-demand read-out. 
We then demonstrated a use-case of our quantum node, as we distributed quantum correlations across a metropolitan quantum network testbed in a multiplexed fashion. We stored signal photons sequentially  in each cell of the QMA and sent the corresponding telecom idler photon to a remote detector station across a $\SI{39.1}{\kilo\meter}$ deployed fiber.
The quantum correlations measured after the long fiber are similar  to the local measurement, indicating the absence of additional noise from the deployed fiber in the detected idlers. 
Finally, we estimated the enhancement in the distribution rate of quantum correlations thanks to the multiplexing capabilities of the node, as we filled up to $\SI{15}{\percent}$ of the communication time with distribution attempts. We showed an increase in detection rate of correlated telecom photons by a factor of 60 compared to a single mode quantum memory. 

We note that in the current experiment, signal photons are retrieved before the detection of the correlated idler at the distant location. A series of hardware improvements, each demonstrated previously in separate systems, will enable efficient long-distance entanglement distribution at enhanced rates provided by our quantum node: Integration of the QMA into an impedance-matched and spatially-multimode cavity will increase memory efficiencies~\cite{Afzelius2010b,Wang2023, Duranti2024, Feldmann2025, Meng2026}.
Millisecond storage times will be achieved through radio-frequency pulses~\cite{Ortu2022a}, applied e.g. through microwave striplines for spatially-selective dynamical decoupling~\cite{Liu2025}.
Stronger filtering of fluorescence noise~\cite{Yang2018} and improved heralding efficiencies of the cSPDC source will increase the quantum correlations of memories and telecom idlers. Larger separations between the memory cells will improve the memory preparation, which will further lower the noise to levels obtained in experiments with single-cell memories~\cite{SuppMat,Haenni2025}. 

\section{Methods}\label{Methods}
\subsection{Evaluation of cross-correlations}
We calculate the cross-correlation $g_{s,i}^{(2)}$ with the probability to detect a coincidence $p_\mathrm{s,i}$ and the unconditional probabilities to detect signal and idler photons, $p_\mathrm{i}$ and $p_\mathrm{s}$. 
The conditional probability $p_\mathrm{s,i}$ is computed by adding up the detection events in a time window $T_\mathrm{DW}$ of a coincidence histogram between signal photons retrieved from the QMA and idler photons.
The unconditional probability $p_\mathrm{s}$ is evaluated on the same time window size $T_\mathrm{DW}$ and at the same position with respect to the CPs, but the histogram is constructed from coincidence events between idler photons and signals of a different pair stored at the next storage trial.
The heralding photon is hence uncorrelated with the signal photon stored in that trial, and the histogram accounts for the noise in the system.
The unconditional probability $p_\mathrm{i}$ is evaluated from the detection rate of idler photons on the same window.

\subsection{Filter elements}
A sequence of optical elements spectrally filters the signal photons retrieved from the QMA: a filter crystal prepared with a transparency window of $\SI{2.4}{\mega\hertz}$, a bandpass filter (Semrock Brightline FF01-600/14), and  an etalon. The SPAD is a COUNT10-FC from Laser Components.

\section*{Acknowledgements}
This project received funding from: EU Horizon Europe Research and Innovation programme (EuroQCI in Spain) (Project no.101091638); Agència de Gestió d'Ajuts Universitaris i de Recerca; Centres de Recerca de Catalunya; FUNDACIÓ Privada MIR-PUIG; Fundación Cellex; Ministerio de Ciencia e Innovación with funding from European Union NextGeneration funds (MCIN/AEI/10.13039/501100011033, PLEC2021-007669 QNetworks, PRTR-C17.I1); Agencia Estatal de Investigación (PID2023-147538OB-I00, Severo Ochoa CEX2024-001490-S); European Union research and innovation program within the Flagship on Quantum Technologies through Horizon Europe project QIA-Phase 1 under grant agreement no. 101102140. European Union’s Horizon 2020 Research and Innovation Programme under the Marie Skłodowska-Curie grant agreement number 956419 (NanoGlass). M.T. acknowledges funding from the European Union's Horizon 2020 research and innovation programme under the Marie Sklodowska-Curie grant agreement No 101103143 ``Two-dimensionally multiplexed on-demand quantum memories'' (2DMultiMems). S.G. acknowledges funding from ``la Caixa'' Foundation (ID 100010434; fellowship code LCF/BQ/PR23/11980044). J.H. acknowledges funding from the “Secretaria d’Universitats i Recerca del Departament de Recerca i Universitats de la Generalitat de Catalunya” under grant 2024 FI-2 00059, as well as the European Social Fund Plus.

\section*{Data availability}
The data of this finding is publicly available in Ref.~\cite{Zenodo2026}
\bibliography{qpsa.bib}


\begin{thebibliography}{43}
\ifx \bisbn   \undefined \def \bisbn  #1{ISBN #1}\fi
\ifx \binits  \undefined \def \binits#1{#1}\fi
\ifx \bauthor  \undefined \def \bauthor#1{#1}\fi
\ifx \batitle  \undefined \def \batitle#1{#1}\fi
\ifx \bjtitle  \undefined \def \bjtitle#1{#1}\fi
\ifx \bvolume  \undefined \def \bvolume#1{\textbf{#1}}\fi
\ifx \byear  \undefined \def \byear#1{#1}\fi
\ifx \bissue  \undefined \def \bissue#1{#1}\fi
\ifx \bfpage  \undefined \def \bfpage#1{#1}\fi
\ifx \blpage  \undefined \def \blpage #1{#1}\fi
\ifx \burl  \undefined \def \burl#1{\textsf{#1}}\fi
\ifx \doiurl  \undefined \def \doiurl#1{\url{https://doi.org/#1}}\fi
\ifx \betal  \undefined \def \betal{\textit{et al.}}\fi
\ifx \binstitute  \undefined \def \binstitute#1{#1}\fi
\ifx \binstitutionaled  \undefined \def \binstitutionaled#1{#1}\fi
\ifx \bctitle  \undefined \def \bctitle#1{#1}\fi
\ifx \beditor  \undefined \def \beditor#1{#1}\fi
\ifx \bpublisher  \undefined \def \bpublisher#1{#1}\fi
\ifx \bbtitle  \undefined \def \bbtitle#1{#1}\fi
\ifx \bedition  \undefined \def \bedition#1{#1}\fi
\ifx \bseriesno  \undefined \def \bseriesno#1{#1}\fi
\ifx \blocation  \undefined \def \blocation#1{#1}\fi
\ifx \bsertitle  \undefined \def \bsertitle#1{#1}\fi
\ifx \bsnm \undefined \def \bsnm#1{#1}\fi
\ifx \bsuffix \undefined \def \bsuffix#1{#1}\fi
\ifx \bparticle \undefined \def \bparticle#1{#1}\fi
\ifx \barticle \undefined \def \barticle#1{#1}\fi
\bibcommenthead
\ifx \bconfdate \undefined \def \bconfdate #1{#1}\fi
\ifx \botherref \undefined \def \botherref #1{#1}\fi
\ifx \url \undefined \def \url#1{\textsf{#1}}\fi
\ifx \bchapter \undefined \def \bchapter#1{#1}\fi
\ifx \bbook \undefined \def \bbook#1{#1}\fi
\ifx \bcomment \undefined \def \bcomment#1{#1}\fi
\ifx \oauthor \undefined \def \oauthor#1{#1}\fi
\ifx \citeauthoryear \undefined \def \citeauthoryear#1{#1}\fi
\ifx \endbibitem  \undefined \def \endbibitem {}\fi
\ifx \bconflocation  \undefined \def \bconflocation#1{#1}\fi
\ifx \arxivurl  \undefined \def \arxivurl#1{\textsf{#1}}\fi
\csname PreBibitemsHook\endcsname

\bibitem[\protect\citeauthoryear{Briegel et~al.}{1998}]{Briegel1998}
\begin{barticle}
\bauthor{\bsnm{Briegel}, \binits{H.-J.}},
\bauthor{\bsnm{D{\"{u}}r}, \binits{W.}},
\bauthor{\bsnm{Cirac}, \binits{J.I.}},
\bauthor{\bsnm{Zoller}, \binits{P.}}:
\batitle{{Quantum Repeaters: The Role of Imperfect Local Operations in Quantum
  Communication}}.
\bjtitle{Phys. Rev. Lett.}
\bvolume{81}(\bissue{26}),
\bfpage{5932}--\blpage{5935}
(\byear{1998})
\doiurl{10.1103/PhysRevLett.81.5932}
\end{barticle}
\endbibitem

\bibitem[\protect\citeauthoryear{Kimble}{2008}]{Kimble2008}
\begin{barticle}
\bauthor{\bsnm{Kimble}, \binits{H.J.}}:
\batitle{{The quantum internet}}.
\bjtitle{Nature}
\bvolume{453}(\bissue{7198}),
\bfpage{1023}--\blpage{1030}
(\byear{2008})
\doiurl{10.1038/nature07127}
\end{barticle}
\endbibitem

\bibitem[\protect\citeauthoryear{Sangouard et~al.}{2011}]{Sangouard2011}
\begin{barticle}
\bauthor{\bsnm{Sangouard}, \binits{N.}},
\bauthor{\bsnm{Simon}, \binits{C.}},
\bauthor{\bsnm{Riedmatten}, \binits{H.}},
\bauthor{\bsnm{Gisin}, \binits{N.}}:
\batitle{Quantum repeaters based on atomic ensembles and linear optics}.
\bjtitle{Rev. Mod. Phys.}
\bvolume{83}(\bissue{1}),
\bfpage{33}--\blpage{80}
(\byear{2011})
\doiurl{10.1103/RevModPhys.83.33}
\end{barticle}
\endbibitem

\bibitem[\protect\citeauthoryear{Wehner et~al.}{2018}]{Wehner2018}
\begin{barticle}
\bauthor{\bsnm{Wehner}, \binits{S.}},
\bauthor{\bsnm{Elkouss}, \binits{D.}},
\bauthor{\bsnm{Hanson}, \binits{R.}}:
\batitle{{Quantum internet: A vision for the road ahead}}.
\bjtitle{Science}
\bvolume{362}(\bissue{6412}),
\bfpage{9288}
(\byear{2018})
\doiurl{10.1126/science.aam9288}
\end{barticle}
\endbibitem

\bibitem[\protect\citeauthoryear{Simon et~al.}{2007}]{Simon2007}
\begin{barticle}
\bauthor{\bsnm{Simon}, \binits{C.}},
\bauthor{\bsnm{Riedmatten}, \binits{H.}},
\bauthor{\bsnm{Afzelius}, \binits{M.}},
\bauthor{\bsnm{Sangouard}, \binits{N.}},
\bauthor{\bsnm{Zbinden}, \binits{H.}},
\bauthor{\bsnm{Gisin}, \binits{N.}}:
\batitle{Quantum repeaters with photon pair sources and multimode memories}.
\bjtitle{Phys. Rev. Lett.}
\bvolume{98},
\bfpage{190503}
(\byear{2007})
\doiurl{10.1103/PhysRevLett.98.190503}
\end{barticle}
\endbibitem

\bibitem[\protect\citeauthoryear{Ortu et~al.}{2022}]{Ortu2022b}
\begin{barticle}
\bauthor{\bsnm{Ortu}, \binits{A.}},
\bauthor{\bsnm{Rakonjac}, \binits{J.V.}},
\bauthor{\bsnm{Holz{\"{a}}pfel}, \binits{A.}},
\bauthor{\bsnm{Seri}, \binits{A.}},
\bauthor{\bsnm{Grandi}, \binits{S.}},
\bauthor{\bsnm{Mazzera}, \binits{M.}},
\bauthor{\bsnm{Riedmatten}, \binits{H.}},
\bauthor{\bsnm{Afzelius}, \binits{M.}}:
\batitle{{Multimode capacity of atomic-frequency comb quantum memories}}.
\bjtitle{Quantum Sci. Technol.}
\bvolume{7}(\bissue{3}),
\bfpage{035024}
(\byear{2022})
\doiurl{10.1088/2058-9565/ac73b0}
\end{barticle}
\endbibitem

\bibitem[\protect\citeauthoryear{Lan et~al.}{2009}]{Lan2009}
\begin{barticle}
\bauthor{\bsnm{Lan}, \binits{S.-Y.}},
\bauthor{\bsnm{Radnaev}, \binits{A.G.}},
\bauthor{\bsnm{Collins}, \binits{O.A.}},
\bauthor{\bsnm{Matsukevich}, \binits{D.N.}},
\bauthor{\bsnm{Kennedy}, \binits{T.A.}},
\bauthor{\bsnm{Kuzmich}, \binits{A.}}:
\batitle{A multiplexed quantum memory}.
\bjtitle{Opt. Express}
\bvolume{17}(\bissue{16}),
\bfpage{13639}--\blpage{13645}
(\byear{2009})
\doiurl{10.1364/OE.17.013639}
\end{barticle}
\endbibitem

\bibitem[\protect\citeauthoryear{Pu et~al.}{2017}]{Pu2017}
\begin{barticle}
\bauthor{\bsnm{Pu}, \binits{Y.-F.}},
\bauthor{\bsnm{Jiang}, \binits{N.}},
\bauthor{\bsnm{Chang}, \binits{W.}},
\bauthor{\bsnm{Yang}, \binits{H.-X.}},
\bauthor{\bsnm{Li}, \binits{C.}},
\bauthor{\bsnm{Duan}, \binits{L.-M.}}:
\batitle{{Experimental realization of a multiplexed quantum memory with 225
  individually accessible memory cells}}.
\bjtitle{Nat. Comm.}
\bvolume{8}(\bissue{1}),
\bfpage{15359}
(\byear{2017})
\doiurl{10.1038/ncomms15359}
\end{barticle}
\endbibitem

\bibitem[\protect\citeauthoryear{Tian et~al.}{2017}]{Tian2017}
\begin{barticle}
\bauthor{\bsnm{Tian}, \binits{L.}},
\bauthor{\bsnm{Xu}, \binits{Z.-X.}},
\bauthor{\bsnm{Chen}, \binits{L.}},
\bauthor{\bsnm{Ge}, \binits{W.}},
\bauthor{\bsnm{Yuan}, \binits{H.}},
\bauthor{\bsnm{Wen}, \binits{Y.-f.}},
\bauthor{\bsnm{Wang}, \binits{S.}},
\bauthor{\bsnm{Li}, \binits{S.}},
\bauthor{\bsnm{Wang}, \binits{H.}}:
\batitle{{Spatial Multiplexing of Atom-Photon Entanglement Sources using
  Feedforward Control and Switching Networks}}.
\bjtitle{Phys. Rev. Lett.}
\bvolume{119}(\bissue{13}),
\bfpage{130505}
(\byear{2017})
\doiurl{10.1103/PhysRevLett.119.130505}
\end{barticle}
\endbibitem

\bibitem[\protect\citeauthoryear{Chang et~al.}{2019}]{Chang2019}
\begin{barticle}
\bauthor{\bsnm{Chang}, \binits{W.}},
\bauthor{\bsnm{Li}, \binits{C.}},
\bauthor{\bsnm{Wu}, \binits{Y.-K.}},
\bauthor{\bsnm{Jiang}, \binits{N.}},
\bauthor{\bsnm{Zhang}, \binits{S.}},
\bauthor{\bsnm{Pu}, \binits{Y.-F.}},
\bauthor{\bsnm{Chang}, \binits{X.-Y.}},
\bauthor{\bsnm{Duan}, \binits{L.-M.}}:
\batitle{{Long-Distance Entanglement between a Multiplexed Quantum Memory and a
  Telecom Photon}}.
\bjtitle{Phys. Rev. X}
\bvolume{9}(\bissue{4}),
\bfpage{041033}
(\byear{2019})
\doiurl{10.1103/PhysRevX.9.041033}
\end{barticle}
\endbibitem

\bibitem[\protect\citeauthoryear{Zhang et~al.}{2024}]{Zhang2024a}
\begin{barticle}
\bauthor{\bsnm{Zhang}, \binits{S.}},
\bauthor{\bsnm{Shi}, \binits{J.}},
\bauthor{\bsnm{Liang}, \binits{Y.}},
\bauthor{\bsnm{Sun}, \binits{Y.}},
\bauthor{\bsnm{Wu}, \binits{Y.}},
\bauthor{\bsnm{Duan}, \binits{L.}},
\bauthor{\bsnm{Pu}, \binits{Y.}}:
\batitle{Fast delivery of heralded atom-photon quantum correlation over 12 km
  fiber through multiplexing enhancement}.
\bjtitle{Nat. Comm.}
\bvolume{15}(\bissue{1}),
\bfpage{10306}
(\byear{2024})
\doiurl{10.1038/s41467-024-54691-3}
\end{barticle}
\endbibitem

\bibitem[\protect\citeauthoryear{Heller et~al.}{2020}]{Heller2020}
\begin{barticle}
\bauthor{\bsnm{Heller}, \binits{L.}},
\bauthor{\bsnm{Farrera}, \binits{P.}},
\bauthor{\bsnm{Heinze}, \binits{G.}},
\bauthor{\bsnm{Riedmatten}, \binits{H.}}:
\batitle{Cold-atom temporally multiplexed quantum memory with cavity-enhanced
  noise suppression}.
\bjtitle{Phys. Rev. Lett.}
\bvolume{124}(\bissue{21}),
\bfpage{210504}
(\byear{2020})
\doiurl{10.1103/PhysRevLett.124.210504}
\end{barticle}
\endbibitem

\bibitem[\protect\citeauthoryear{Afzelius et~al.}{2009}]{Afzelius2009}
\begin{barticle}
\bauthor{\bsnm{Afzelius}, \binits{M.}},
\bauthor{\bsnm{Simon}, \binits{C.}},
\bauthor{\bsnm{Riedmatten}, \binits{H.}},
\bauthor{\bsnm{Gisin}, \binits{N.}}:
\batitle{Multimode quantum memory based on atomic frequency combs}.
\bjtitle{Phys. Rev. A}
\bvolume{79},
\bfpage{052329}
(\byear{2009})
\doiurl{10.1103/PhysRevA.79.052329}
\end{barticle}
\endbibitem

\bibitem[\protect\citeauthoryear{Lago-Rivera et~al.}{2021}]{Lago-Rivera2021}
\begin{barticle}
\bauthor{\bsnm{Lago-Rivera}, \binits{D.}},
\bauthor{\bsnm{Grandi}, \binits{S.}},
\bauthor{\bsnm{Rakonjac}, \binits{J.V.}},
\bauthor{\bsnm{Seri}, \binits{A.}},
\bauthor{\bsnm{Riedmatten}, \binits{H.}}:
\batitle{{Telecom-heralded entanglement between multimode solid-state quantum
  memories}}.
\bjtitle{Nature}
\bvolume{594}(\bissue{7861}),
\bfpage{37}--\blpage{40}
(\byear{2021})
\doiurl{10.1038/s41586-021-03481-8}
\end{barticle}
\endbibitem

\bibitem[\protect\citeauthoryear{Businger et~al.}{2022}]{Businger2022}
\begin{barticle}
\bauthor{\bsnm{Businger}, \binits{M.}},
\bauthor{\bsnm{Nicolas}, \binits{L.}},
\bauthor{\bsnm{Mejia}, \binits{T.S.}},
\bauthor{\bsnm{Ferrier}, \binits{A.}},
\bauthor{\bsnm{Goldner}, \binits{P.}},
\bauthor{\bsnm{Afzelius}, \binits{M.}}:
\batitle{{Non-classical correlations over 1250 modes between telecom photons
  and 979-nm photons stored in $^{171}$Yb$^{3+}$:Y$_2$SiO$_5$}}.
\bjtitle{Nature Communications}
\bvolume{13}(\bissue{1}),
\bfpage{6438}
(\byear{2022})
\doiurl{10.1038/s41467-022-33929-y}
\end{barticle}
\endbibitem

\bibitem[\protect\citeauthoryear{Zhu et~al.}{2026}]{Zhu2026}
\begin{barticle}
\bauthor{\bsnm{Zhu}, \binits{T.-X.}},
\bauthor{\bsnm{Zhang}, \binits{C.}},
\bauthor{\bsnm{Ou}, \binits{Z.-W.}},
\bauthor{\bsnm{Liu}, \binits{X.}},
\bauthor{\bsnm{Liang}, \binits{P.-J.}},
\bauthor{\bsnm{Hu}, \binits{X.-M.}},
\bauthor{\bsnm{Huang}, \binits{Y.-F.}},
\bauthor{\bsnm{Zhou}, \binits{Z.-Q.}},
\bauthor{\bsnm{Li}, \binits{C.-F.}},
\bauthor{\bsnm{Guo}, \binits{G.-C.}}:
\batitle{A metropolitan-scale multiplexed quantum repeater with bell
  non-locality}.
\bjtitle{Nature Photonics}
(\byear{2026})
\doiurl{10.1038/s41566-026-01911-5}
\end{barticle}
\endbibitem

\bibitem[\protect\citeauthoryear{G\"{u}ndo\u{g}an et~al.}{2012}]{Gundogan2012}
\begin{barticle}
\bauthor{\bsnm{G\"{u}ndo\u{g}an}, \binits{M.}},
\bauthor{\bsnm{Ledingham}, \binits{P.M.}},
\bauthor{\bsnm{Almasi}, \binits{A.}},
\bauthor{\bsnm{Cristiani}, \binits{M.}},
\bauthor{\bsnm{Riedmatten}, \binits{H.}}:
\batitle{Quantum storage of a photonic polarization qubit in a solid}.
\bjtitle{Phys. Rev. Lett.}
\bvolume{108}(\bissue{19}),
\bfpage{190504}
(\byear{2012})
\doiurl{10.1103/PhysRevLett.108.190504}
\end{barticle}
\endbibitem

\bibitem[\protect\citeauthoryear{Yang et~al.}{2018}]{Yang2018}
\begin{barticle}
\bauthor{\bsnm{Yang}, \binits{T.-S.}},
\bauthor{\bsnm{Zhou}, \binits{Z.-Q.}},
\bauthor{\bsnm{Hua}, \binits{Y.-L.}},
\bauthor{\bsnm{Liu}, \binits{X.}},
\bauthor{\bsnm{Li}, \binits{Z.-F.}},
\bauthor{\bsnm{Li}, \binits{P.-Y.}},
\bauthor{\bsnm{Ma}, \binits{Y.}},
\bauthor{\bsnm{Liu}, \binits{C.}},
\bauthor{\bsnm{Liang}, \binits{P.-J.}},
\bauthor{\bsnm{Li}, \binits{X.}},
\bauthor{\bsnm{Xiao}, \binits{Y.-X.}},
\bauthor{\bsnm{Hu}, \binits{J.}},
\bauthor{\bsnm{Li}, \binits{C.-F.}},
\bauthor{\bsnm{Guo}, \binits{G.-C.}}:
\batitle{{Multiplexed storage and real-time manipulation based on a multiple
  degree-of-freedom quantum memory}}.
\bjtitle{Nat. Comm.}
\bvolume{9}(\bissue{1}),
\bfpage{3407}
(\byear{2018})
\doiurl{10.1038/s41467-018-05669-5}
\end{barticle}
\endbibitem

\bibitem[\protect\citeauthoryear{Teller et~al.}{2025}]{Teller2025a}
\begin{barticle}
\bauthor{\bsnm{Teller}, \binits{M.}},
\bauthor{\bsnm{Plascencia}, \binits{S.}},
\bauthor{\bsnm{Sastre~Jachimska}, \binits{C.}},
\bauthor{\bsnm{Grandi}, \binits{S.}},
\bauthor{\bsnm{Riedmatten}, \binits{H.}}:
\batitle{A solid-state temporally multiplexed quantum memory array at the
  single-photon level}.
\bjtitle{npj Quantum Information}
\bvolume{11}(\bissue{1}),
\bfpage{92}
(\byear{2025})
\doiurl{10.1038/s41534-025-01042-9}
\end{barticle}
\endbibitem

\bibitem[\protect\citeauthoryear{Ou et~al.}{2025}]{Ou2025}
\begin{botherref}
\oauthor{\bsnm{Ou}, \binits{Z.-W.}},
\oauthor{\bsnm{Zhu}, \binits{T.-X.}},
\oauthor{\bsnm{Liang}, \binits{P.-J.}},
\oauthor{\bsnm{Hu}, \binits{X.-M.}},
\oauthor{\bsnm{Zhou}, \binits{Z.-Q.}},
\oauthor{\bsnm{Li}, \binits{C.-F.}},
\oauthor{\bsnm{Guo}, \binits{G.-C.}}:
Multichannel and high dimensional integrated photonic quantum memory
(2025).
\url{https://arxiv.org/abs/2508.19605}
\end{botherref}
\endbibitem

\bibitem[\protect\citeauthoryear{Sinclair et~al.}{2014}]{Sinclair2014}
\begin{barticle}
\bauthor{\bsnm{Sinclair}, \binits{N.}},
\bauthor{\bsnm{Saglamyurek}, \binits{E.}},
\bauthor{\bsnm{Mallahzadeh}, \binits{H.}},
\bauthor{\bsnm{Slater}, \binits{J.A.}},
\bauthor{\bsnm{George}, \binits{M.}},
\bauthor{\bsnm{Ricken}, \binits{R.}},
\bauthor{\bsnm{Hedges}, \binits{M.P.}},
\bauthor{\bsnm{Oblak}, \binits{D.}},
\bauthor{\bsnm{Simon}, \binits{C.}},
\bauthor{\bsnm{Sohler}, \binits{W.}},
\bauthor{\bsnm{Tittel}, \binits{W.}}:
\batitle{Spectral multiplexing for scalable quantum photonics using an atomic
  frequency comb quantum memory and feed-forward control}.
\bjtitle{Phys. Rev. Lett.}
\bvolume{113}(\bissue{5}),
\bfpage{053603}
(\byear{2014})
\doiurl{10.1103/PhysRevLett.113.053603}
\end{barticle}
\endbibitem

\bibitem[\protect\citeauthoryear{Seri et~al.}{2019}]{Seri2019}
\begin{barticle}
\bauthor{\bsnm{Seri}, \binits{A.}},
\bauthor{\bsnm{Lago-Rivera}, \binits{D.}},
\bauthor{\bsnm{Lenhard}, \binits{A.}},
\bauthor{\bsnm{Corrielli}, \binits{G.}},
\bauthor{\bsnm{Osellame}, \binits{R.}},
\bauthor{\bsnm{Mazzera}, \binits{M.}},
\bauthor{\bsnm{Riedmatten}, \binits{H.}}:
\batitle{{Quantum Storage of Frequency-Multiplexed Heralded Single Photons}}.
\bjtitle{Phys. Rev. Lett.}
\bvolume{123}(\bissue{8}),
\bfpage{080502}
(\byear{2019})
\doiurl{10.1103/PhysRevLett.123.080502}
\end{barticle}
\endbibitem

\bibitem[\protect\citeauthoryear{Tateishi et~al.}{2026}]{Tateishi2026}
\begin{botherref}
\oauthor{\bsnm{Tateishi}, \binits{H.}},
\oauthor{\bsnm{Yoshida}, \binits{D.}},
\oauthor{\bsnm{Tsuno}, \binits{T.}},
\oauthor{\bsnm{Nihashi}, \binits{T.}},
\oauthor{\bsnm{Komatsudaira}, \binits{R.}},
\oauthor{\bsnm{Akamatsu}, \binits{D.}},
\oauthor{\bsnm{Hong}, \binits{F.-L.}},
\oauthor{\bsnm{Nagano}, \binits{K.}},
\oauthor{\bsnm{Horikiri}, \binits{T.}}:
Quantum storage of frequency-multiplexed photons exhibiting nonclassical
  correlations with telecom c-band photons.
Applied Physics Letters
\textbf{128}(21)
(2026)
\doiurl{10.1063/5.0313240}
\end{botherref}
\endbibitem

\bibitem[\protect\citeauthoryear{Fekete et~al.}{2013}]{Fekete2013}
\begin{barticle}
\bauthor{\bsnm{Fekete}, \binits{J.}},
\bauthor{\bsnm{Riel\"ander}, \binits{D.}},
\bauthor{\bsnm{Cristiani}, \binits{M.}},
\bauthor{\bsnm{Riedmatten}, \binits{H.}}:
\batitle{Ultranarrow-band photon-pair source compatible with solid state
  quantum memories and telecommunication networks}.
\bjtitle{Phys. Rev. Lett.}
\bvolume{110},
\bfpage{220502}
(\byear{2013})
\doiurl{10.1103/PhysRevLett.110.220502}
\end{barticle}
\endbibitem

\bibitem[\protect\citeauthoryear{Riel{\"{a}}nder et~al.}{2016}]{Rielander2016}
\begin{barticle}
\bauthor{\bsnm{Riel{\"{a}}nder}, \binits{D.}},
\bauthor{\bsnm{Lenhard}, \binits{A.}},
\bauthor{\bsnm{Mazzera}, \binits{M.}},
\bauthor{\bsnm{Riedmatten}, \binits{H.}}:
\batitle{Cavity enhanced telecom heralded single photons for spin-wave solid
  state quantum memories}.
\bjtitle{New J. Phys.}
\bvolume{18}(\bissue{12}),
\bfpage{123013}
(\byear{2016})
\doiurl{10.1088/1367-2630/aa4f38}
\end{barticle}
\endbibitem

\bibitem[\protect\citeauthoryear{Teller et~al.}{2025}]{Teller2025}
\begin{barticle}
\bauthor{\bsnm{Teller}, \binits{M.}},
\bauthor{\bsnm{Plascencia}, \binits{S.}},
\bauthor{\bsnm{Grandi}, \binits{S.}},
\bauthor{\bsnm{Riedmatten}, \binits{H.}}:
\batitle{Quantum storage of qubits in an array of independently controllable
  solid-state quantum memories}.
\bjtitle{Phys. Rev. X}
\bvolume{15},
\bfpage{031053}
(\byear{2025})
\doiurl{10.1103/z6lc-qw2d}
\end{barticle}
\endbibitem

\bibitem[\protect\citeauthoryear{Rakonjac et~al.}{2023}]{Rakonjac2023}
\begin{barticle}
\bauthor{\bsnm{Rakonjac}, \binits{J.V.}},
\bauthor{\bsnm{Grandi}, \binits{S.}},
\bauthor{\bsnm{Wengerowsky}, \binits{S.}},
\bauthor{\bsnm{Lago-Rivera}, \binits{D.}},
\bauthor{\bsnm{Appas}, \binits{F.}},
\bauthor{\bsnm{Riedmatten}, \binits{H.}}:
\batitle{Transmission of light--matter entanglement over a metropolitan
  network}.
\bjtitle{Optica Quantum}
\bvolume{1}(\bissue{2}),
\bfpage{94}--\blpage{102}
(\byear{2023})
\doiurl{10.1364/OPTICAQ.501048}
\end{barticle}
\endbibitem

\bibitem[\protect\citeauthoryear{Stolk et~al.}{2024}]{Stolk2024}
\begin{botherref}
\oauthor{\bsnm{Stolk}, \binits{A.J.}},
\oauthor{\bsnm{Enden}, \binits{K.L.}},
\oauthor{\bsnm{Slater}, \binits{M.-C.}},
\oauthor{\bsnm{Raa-Derckx}, \binits{I.}},
\oauthor{\bsnm{Botma}, \binits{P.}},
\oauthor{\bsnm{Rantwijk}, \binits{J.}},
\oauthor{\bsnm{Biemond}, \binits{J.J.B.}},
\oauthor{\bsnm{Hagen}, \binits{R.A.J.}},
\oauthor{\bsnm{Herfst}, \binits{R.W.}},
\oauthor{\bsnm{Koek}, \binits{W.D.}},
\oauthor{\bsnm{Meskers}, \binits{A.J.H.}},
\oauthor{\bsnm{Vollmer}, \binits{R.}},
\oauthor{\bsnm{Zwet}, \binits{E.J.}},
\oauthor{\bsnm{Markham}, \binits{M.}},
\oauthor{\bsnm{Edmonds}, \binits{A.M.}},
\oauthor{\bsnm{Geus}, \binits{J.F.}},
\oauthor{\bsnm{Elsen}, \binits{F.}},
\oauthor{\bsnm{Jungbluth}, \binits{B.}},
\oauthor{\bsnm{Haefner}, \binits{C.}},
\oauthor{\bsnm{Tresp}, \binits{C.}},
\oauthor{\bsnm{Stuhler}, \binits{J.}},
\oauthor{\bsnm{Ritter}, \binits{S.}},
\oauthor{\bsnm{Hanson}, \binits{R.}}:
Metropolitan-scale heralded entanglement of solid-state qubits.
Science Advances
\textbf{10}(44)
(2024)
\doiurl{10.1126/sciadv.adp6442}
\end{botherref}
\endbibitem

\bibitem[\protect\citeauthoryear{Knaut et~al.}{2024}]{Knaut2024}
\begin{barticle}
\bauthor{\bsnm{Knaut}, \binits{C.M.}},
\bauthor{\bsnm{Suleymanzade}, \binits{A.}},
\bauthor{\bsnm{Wei}, \binits{Y.-C.}},
\bauthor{\bsnm{Assumpcao}, \binits{D.R.}},
\bauthor{\bsnm{Stas}, \binits{P.-J.}},
\bauthor{\bsnm{Huan}, \binits{Y.Q.}},
\bauthor{\bsnm{Machielse}, \binits{B.}},
\bauthor{\bsnm{Knall}, \binits{E.N.}},
\bauthor{\bsnm{Sutula}, \binits{M.}},
\bauthor{\bsnm{Baranes}, \binits{G.}},
\bauthor{\bsnm{Sinclair}, \binits{N.}},
\bauthor{\bsnm{De-Eknamkul}, \binits{C.}},
\bauthor{\bsnm{Levonian}, \binits{D.S.}},
\bauthor{\bsnm{Bhaskar}, \binits{M.K.}},
\bauthor{\bsnm{Park}, \binits{H.}},
\bauthor{\bsnm{Lončar}, \binits{M.}},
\bauthor{\bsnm{Lukin}, \binits{M.D.}}:
\batitle{Entanglement of nanophotonic quantum memory nodes in a telecom
  network}.
\bjtitle{Nature}
\bvolume{629}(\bissue{8012}),
\bfpage{573}--\blpage{578}
(\byear{2024})
\doiurl{10.1038/s41586-024-07252-z}
\end{barticle}
\endbibitem

\bibitem[\protect\citeauthoryear{Liu et~al.}{2024}]{Liu2024}
\begin{barticle}
\bauthor{\bsnm{Liu}, \binits{J.-L.}},
\bauthor{\bsnm{Luo}, \binits{X.-Y.}},
\bauthor{\bsnm{Yu}, \binits{Y.}},
\bauthor{\bsnm{Wang}, \binits{C.-Y.}},
\bauthor{\bsnm{Wang}, \binits{B.}},
\bauthor{\bsnm{Hu}, \binits{Y.}},
\bauthor{\bsnm{Li}, \binits{J.}},
\bauthor{\bsnm{Zheng}, \binits{M.-Y.}},
\bauthor{\bsnm{Yao}, \binits{B.}},
\bauthor{\bsnm{Yan}, \binits{Z.}},
\bauthor{\bsnm{Teng}, \binits{D.}},
\bauthor{\bsnm{Jiang}, \binits{J.-W.}},
\bauthor{\bsnm{Liu}, \binits{X.-B.}},
\bauthor{\bsnm{Xie}, \binits{X.-P.}},
\bauthor{\bsnm{Zhang}, \binits{J.}},
\bauthor{\bsnm{Mao}, \binits{Q.-H.}},
\bauthor{\bsnm{Jiang}, \binits{X.}},
\bauthor{\bsnm{Zhang}, \binits{Q.}},
\bauthor{\bsnm{Bao}, \binits{X.-H.}},
\bauthor{\bsnm{Pan}, \binits{J.-W.}}:
\batitle{Creation of memory–memory entanglement in a metropolitan quantum
  network}.
\bjtitle{Nature}
\bvolume{629}(\bissue{8012}),
\bfpage{579}--\blpage{585}
(\byear{2024})
\doiurl{10.1038/s41586-024-07308-0}
\end{barticle}
\endbibitem

\bibitem[\protect\citeauthoryear{Afzelius et~al.}{2010}]{Afzelius2010a}
\begin{barticle}
\bauthor{\bsnm{Afzelius}, \binits{M.}},
\bauthor{\bsnm{Usmani}, \binits{I.}},
\bauthor{\bsnm{Amari}, \binits{B.} \bsuffix{A.and~Lauritzen}},
\bauthor{\bsnm{Walther}, \binits{A.}},
\bauthor{\bsnm{Simon}, \binits{C.}},
\bauthor{\bsnm{Sangouard}, \binits{N.}},
\bauthor{\bsnm{Min\'ar}, \binits{J.}},
\bauthor{\bsnm{Riedmatten}, \binits{H.}},
\bauthor{\bsnm{Gisin}, \binits{N.}},
\bauthor{\bsnm{Kr\"oll}, \binits{S.}}:
\batitle{Demonstration of atomic frequency comb memory for light with spin-wave
  storage}.
\bjtitle{Phys. Rev. Lett.}
\bvolume{104}(\bissue{4}),
\bfpage{040503}
(\byear{2010})
\doiurl{10.1103/PhysRevLett.104.040503}
\end{barticle}
\endbibitem

\bibitem[\protect\citeauthoryear{Seri et~al.}{2017}]{Seri2017}
\begin{barticle}
\bauthor{\bsnm{Seri}, \binits{A.}},
\bauthor{\bsnm{Lenhard}, \binits{A.}},
\bauthor{\bsnm{Riel{\"{a}}nder}, \binits{D.}},
\bauthor{\bsnm{G{\"{u}}ndoğan}, \binits{M.}},
\bauthor{\bsnm{Ledingham}, \binits{P.M.}},
\bauthor{\bsnm{Mazzera}, \binits{M.}},
\bauthor{\bsnm{Riedmatten}, \binits{H.}}:
\batitle{{Quantum Correlations between Single Telecom Photons and a Multimode
  On-Demand Solid-State Quantum Memory}}.
\bjtitle{Physical Review X}
\bvolume{7}(\bissue{2}),
\bfpage{021028}
(\byear{2017})
\doiurl{10.1103/PhysRevX.7.021028}
\end{barticle}
\endbibitem

\bibitem[\protect\citeauthoryear{Delle~Donne et~al.}{2025}]{DelleDonne2025}
\begin{barticle}
\bauthor{\bsnm{Delle~Donne}, \binits{C.}},
\bauthor{\bsnm{Iuliano}, \binits{M.}},
\bauthor{\bsnm{Vecht}, \binits{B.}},
\bauthor{\bsnm{Ferreira}, \binits{G.M.}},
\bauthor{\bsnm{Jirovsk{\'a}}, \binits{H.}},
\bauthor{\bsnm{Steenhoven}, \binits{T.J.W.}},
\bauthor{\bsnm{Dahlberg}, \binits{A.}},
\bauthor{\bsnm{Skrzypczyk}, \binits{M.}},
\bauthor{\bsnm{Fioretto}, \binits{D.}},
\bauthor{\bsnm{Teller}, \binits{M.}},
\bauthor{\bsnm{Filippov}, \binits{P.}},
\bauthor{\bsnm{Montblanch}, \binits{A.R.-P.}},
\bauthor{\bsnm{Fischer}, \binits{J.}},
\bauthor{\bsnm{Ommen}, \binits{H.B.}},
\bauthor{\bsnm{Demetriou}, \binits{N.}},
\bauthor{\bsnm{Leichtle}, \binits{D.}},
\bauthor{\bsnm{Music}, \binits{L.}},
\bauthor{\bsnm{Ollivier}, \binits{H.}},
\bauthor{\bsnm{Raa}, \binits{I.}},
\bauthor{\bsnm{Kozlowski}, \binits{W.}},
\bauthor{\bsnm{Taminiau}, \binits{T.H.}},
\bauthor{\bsnm{Pawe{\l}czak}, \binits{P.}},
\bauthor{\bsnm{Northup}, \binits{T.E.}},
\bauthor{\bsnm{Hanson}, \binits{R.}},
\bauthor{\bsnm{Wehner}, \binits{S.}}:
\batitle{An operating system for executing applications on quantum network
  nodes}.
\bjtitle{Nature}
\bvolume{639}(\bissue{8054}),
\bfpage{321}--\blpage{328}
(\byear{2025})
\doiurl{10.1038/s41586-025-08704-w}
\end{barticle}
\endbibitem

\bibitem[\protect\citeauthoryear{H\"anni et~al.}{2025}]{Haenni2025}
\begin{barticle}
\bauthor{\bsnm{H\"anni}, \binits{J.}},
\bauthor{\bsnm{Rodr\'{\i}guez-Moldes}, \binits{A.E.}},
\bauthor{\bsnm{Appas}, \binits{F.}},
\bauthor{\bsnm{Wengerowsky}, \binits{S.}},
\bauthor{\bsnm{Lago-Rivera}, \binits{D.}},
\bauthor{\bsnm{Teller}, \binits{M.}},
\bauthor{\bsnm{Grandi}, \binits{S.}},
\bauthor{\bsnm{Riedmatten}, \binits{H.}}:
\batitle{Heralded entanglement of on-demand spin-wave solid-state quantum
  memories for multiplexed quantum network links}.
\bjtitle{Phys. Rev. X}
\bvolume{15},
\bfpage{041003}
(\byear{2025})
\doiurl{10.1103/wvv1-6lg8}
\end{barticle}
\endbibitem

\bibitem[\protect\citeauthoryear{Afzelius and Simon}{2010}]{Afzelius2010b}
\begin{barticle}
\bauthor{\bsnm{Afzelius}, \binits{M.}},
\bauthor{\bsnm{Simon}, \binits{C.}}:
\batitle{Impedance-matched cavity quantum memory}.
\bjtitle{Phys. Rev. A}
\bvolume{82}(\bissue{2}),
\bfpage{022310}
(\byear{2010})
\doiurl{10.1103/PhysRevA.82.022310}
\end{barticle}
\endbibitem

\bibitem[\protect\citeauthoryear{Wang et~al.}{2023}]{Wang2023}
\begin{botherref}
\oauthor{\bsnm{Wang}, \binits{M.}},
\oauthor{\bsnm{Jiao}, \binits{H.}},
\oauthor{\bsnm{Lu}, \binits{J.}},
\oauthor{\bsnm{Fan}, \binits{W.}},
\oauthor{\bsnm{Yang}, \binits{Z.}},
\oauthor{\bsnm{Xi}, \binits{M.}},
\oauthor{\bsnm{Li}, \binits{S.}},
\oauthor{\bsnm{Wang}, \binits{H.}}:
Cavity-enhanced and spatial-multimode spin-wave-photon quantum interface
(2023).
\url{https://arxiv.org/abs/2307.12523}
\end{botherref}
\endbibitem

\bibitem[\protect\citeauthoryear{Duranti et~al.}{2024}]{Duranti2024}
\begin{barticle}
\bauthor{\bsnm{Duranti}, \binits{S.}},
\bauthor{\bsnm{Wengerowsky}, \binits{S.}},
\bauthor{\bsnm{Feldmann}, \binits{L.}},
\bauthor{\bsnm{Seri}, \binits{A.}},
\bauthor{\bsnm{Casabone}, \binits{B.}},
\bauthor{\bsnm{Riedmatten}, \binits{H.}}:
\batitle{Efficient cavity-assisted storage of photonic qubits in a solid-state
  quantum memory}.
\bjtitle{Opt. Express}
\bvolume{32}(\bissue{15}),
\bfpage{26884}--\blpage{26895}
(\byear{2024})
\doiurl{10.1364/OE.512318}
\end{barticle}
\endbibitem

\bibitem[\protect\citeauthoryear{Feldmann et~al.}{2025}]{Feldmann2025}
\begin{botherref}
\oauthor{\bsnm{Feldmann}, \binits{L.}},
\oauthor{\bsnm{Wengerowsky}, \binits{S.}},
\oauthor{\bsnm{Das}, \binits{A.}},
\oauthor{\bsnm{Duranti}, \binits{S.}},
\oauthor{\bsnm{H\"{a}nni}, \binits{J.}},
\oauthor{\bsnm{Grandi}, \binits{S.}},
\oauthor{\bsnm{Riedmatten}, \binits{H.}}:
Cavity-enhanced spin-wave solid-state quantum memory.
Physical Review Letters
\textbf{135}(12)
(2025)
\doiurl{10.1103/8l9k-12k2}
\end{botherref}
\endbibitem

\bibitem[\protect\citeauthoryear{Meng et~al.}{2026}]{Meng2026}
\begin{barticle}
\bauthor{\bsnm{Meng}, \binits{R.-R.}},
\bauthor{\bsnm{Liu}, \binits{P.-X.}},
\bauthor{\bsnm{Liu}, \binits{X.}},
\bauthor{\bsnm{Zhu}, \binits{T.-X.}},
\bauthor{\bsnm{Liang}, \binits{P.-J.}},
\bauthor{\bsnm{Zhang}, \binits{C.}},
\bauthor{\bsnm{Tang}, \binits{Z.-Y.}},
\bauthor{\bsnm{Zhang}, \binits{H.-Z.}},
\bauthor{\bsnm{Cui}, \binits{J.-M.}},
\bauthor{\bsnm{Jin}, \binits{M.}},
\bauthor{\bsnm{Zhou}, \binits{Z.-Q.}},
\bauthor{\bsnm{Li}, \binits{C.-F.}},
\bauthor{\bsnm{Guo}, \binits{G.-C.}}:
\batitle{Efficient integrated quantum memory for light}.
\bjtitle{Nature Photonics}
\bvolume{20}(\bissue{4}),
\bfpage{437}--\blpage{443}
(\byear{2026})
\doiurl{10.1038/s41566-026-01845-y}
\end{barticle}
\endbibitem

\bibitem[\protect\citeauthoryear{Ortu et~al.}{2022}]{Ortu2022a}
\begin{barticle}
\bauthor{\bsnm{Ortu}, \binits{A.}},
\bauthor{\bsnm{Holz{\"{a}}pfel}, \binits{A.}},
\bauthor{\bsnm{Etesse}, \binits{J.}},
\bauthor{\bsnm{Afzelius}, \binits{M.}}:
\batitle{{Storage of photonic time-bin qubits for up to 20 ms in a rare-earth
  doped crystal}}.
\bjtitle{Npj Quantum Inf.}
\bvolume{8}(\bissue{1}),
\bfpage{29}
(\byear{2022})
\doiurl{10.1038/s41534-022-00541-3}
\end{barticle}
\endbibitem

\bibitem[\protect\citeauthoryear{Liu et~al.}{2025}]{Liu2025}
\begin{barticle}
\bauthor{\bsnm{Liu}, \binits{Y.-P.}},
\bauthor{\bsnm{Ou}, \binits{Z.-W.}},
\bauthor{\bsnm{Zhu}, \binits{T.-X.}},
\bauthor{\bsnm{Su}, \binits{M.-X.}},
\bauthor{\bsnm{Liu}, \binits{C.}},
\bauthor{\bsnm{Han}, \binits{Y.-J.}},
\bauthor{\bsnm{Zhou}, \binits{Z.-Q.}},
\bauthor{\bsnm{Li}, \binits{C.-F.}},
\bauthor{\bsnm{Guo}, \binits{G.-C.}}:
\batitle{A millisecond integrated quantum memory for photonic qubits}.
\bjtitle{Science Advances}
\bvolume{11}(\bissue{13}),
\bfpage{5264}
(\byear{2025})
\doiurl{10.1126/sciadv.adu5264} .
Accessed 2025-03-31
\end{barticle}
\endbibitem

\bibitem[\protect\citeauthoryear{}{}]{SuppMat}
\begin{botherref}
See Supplemental Material at [URL will be inserted by publisher] for [give
  brief description of material].
\end{botherref}
\endbibitem

\bibitem[\protect\citeauthoryear{}{}]{Zenodo2026}
\begin{botherref}
\url{https://doi.org/10.5281/zenodo.20527400}
\end{botherref}
\endbibitem

\end{thebibliography}



\begin{thebibliography}{6}
\ifx \bisbn   \undefined \def \bisbn  #1{ISBN #1}\fi
\ifx \binits  \undefined \def \binits#1{#1}\fi
\ifx \bauthor  \undefined \def \bauthor#1{#1}\fi
\ifx \batitle  \undefined \def \batitle#1{#1}\fi
\ifx \bjtitle  \undefined \def \bjtitle#1{#1}\fi
\ifx \bvolume  \undefined \def \bvolume#1{\textbf{#1}}\fi
\ifx \byear  \undefined \def \byear#1{#1}\fi
\ifx \bissue  \undefined \def \bissue#1{#1}\fi
\ifx \bfpage  \undefined \def \bfpage#1{#1}\fi
\ifx \blpage  \undefined \def \blpage #1{#1}\fi
\ifx \burl  \undefined \def \burl#1{\textsf{#1}}\fi
\ifx \doiurl  \undefined \def \doiurl#1{\url{https://doi.org/#1}}\fi
\ifx \betal  \undefined \def \betal{\textit{et al.}}\fi
\ifx \binstitute  \undefined \def \binstitute#1{#1}\fi
\ifx \binstitutionaled  \undefined \def \binstitutionaled#1{#1}\fi
\ifx \bctitle  \undefined \def \bctitle#1{#1}\fi
\ifx \beditor  \undefined \def \beditor#1{#1}\fi
\ifx \bpublisher  \undefined \def \bpublisher#1{#1}\fi
\ifx \bbtitle  \undefined \def \bbtitle#1{#1}\fi
\ifx \bedition  \undefined \def \bedition#1{#1}\fi
\ifx \bseriesno  \undefined \def \bseriesno#1{#1}\fi
\ifx \blocation  \undefined \def \blocation#1{#1}\fi
\ifx \bsertitle  \undefined \def \bsertitle#1{#1}\fi
\ifx \bsnm \undefined \def \bsnm#1{#1}\fi
\ifx \bsuffix \undefined \def \bsuffix#1{#1}\fi
\ifx \bparticle \undefined \def \bparticle#1{#1}\fi
\ifx \barticle \undefined \def \barticle#1{#1}\fi
\bibcommenthead
\ifx \bconfdate \undefined \def \bconfdate #1{#1}\fi
\ifx \botherref \undefined \def \botherref #1{#1}\fi
\ifx \url \undefined \def \url#1{\textsf{#1}}\fi
\ifx \bchapter \undefined \def \bchapter#1{#1}\fi
\ifx \bbook \undefined \def \bbook#1{#1}\fi
\ifx \bcomment \undefined \def \bcomment#1{#1}\fi
\ifx \oauthor \undefined \def \oauthor#1{#1}\fi
\ifx \citeauthoryear \undefined \def \citeauthoryear#1{#1}\fi
\ifx \endbibitem  \undefined \def \endbibitem {}\fi
\ifx \bconflocation  \undefined \def \bconflocation#1{#1}\fi
\ifx \arxivurl  \undefined \def \arxivurl#1{\textsf{#1}}\fi
\csname PreBibitemsHook\endcsname

\bibitem[\protect\citeauthoryear{Albrecht et~al.}{2014}]{Albrecht2014}
\begin{barticle}
\bauthor{\bsnm{Albrecht}, \binits{B.}},
\bauthor{\bsnm{Farrera}, \binits{P.}},
\bauthor{\bsnm{Fernandez-Gonzalvo}, \binits{X.}},
\bauthor{\bsnm{Cristiani}, \binits{M.}},
\bauthor{\bsnm{Riedmatten}, \binits{H.}}:
\batitle{A waveguide frequency converter connecting rubidium-based quantum
  memories to the telecom c-band}.
\bjtitle{Nat. Comm.}
\bvolume{5},
\bfpage{3376}
(\byear{2014})
\doiurl{10.1038/ncomms43760}
\end{barticle}
\endbibitem

\bibitem[\protect\citeauthoryear{Teller et~al.}{2025}]{Teller2025a}
\begin{barticle}
\bauthor{\bsnm{Teller}, \binits{M.}},
\bauthor{\bsnm{Plascencia}, \binits{S.}},
\bauthor{\bsnm{Sastre~Jachimska}, \binits{C.}},
\bauthor{\bsnm{Grandi}, \binits{S.}},
\bauthor{\bsnm{Riedmatten}, \binits{H.}}:
\batitle{A solid-state temporally multiplexed quantum memory array at the
  single-photon level}.
\bjtitle{npj Quantum Information}
\bvolume{11}(\bissue{1}),
\bfpage{92}
(\byear{2025})
\doiurl{10.1038/s41534-025-01042-9}
\end{barticle}
\endbibitem

\bibitem[\protect\citeauthoryear{H\"anni et~al.}{2025}]{Haenni2025}
\begin{barticle}
\bauthor{\bsnm{H\"anni}, \binits{J.}},
\bauthor{\bsnm{Rodr\'{\i}guez-Moldes}, \binits{A.E.}},
\bauthor{\bsnm{Appas}, \binits{F.}},
\bauthor{\bsnm{Wengerowsky}, \binits{S.}},
\bauthor{\bsnm{Lago-Rivera}, \binits{D.}},
\bauthor{\bsnm{Teller}, \binits{M.}},
\bauthor{\bsnm{Grandi}, \binits{S.}},
\bauthor{\bsnm{Riedmatten}, \binits{H.}}:
\batitle{Heralded entanglement of on-demand spin-wave solid-state quantum
  memories for multiplexed quantum network links}.
\bjtitle{Phys. Rev. X}
\bvolume{15},
\bfpage{041003}
(\byear{2025})
\doiurl{10.1103/wvv1-6lg8}
\end{barticle}
\endbibitem

\bibitem[\protect\citeauthoryear{Fekete et~al.}{2013}]{Fekete2013}
\begin{barticle}
\bauthor{\bsnm{Fekete}, \binits{J.}},
\bauthor{\bsnm{Riel\"ander}, \binits{D.}},
\bauthor{\bsnm{Cristiani}, \binits{M.}},
\bauthor{\bsnm{Riedmatten}, \binits{H.}}:
\batitle{Ultranarrow-band photon-pair source compatible with solid state
  quantum memories and telecommunication networks}.
\bjtitle{Phys. Rev. Lett.}
\bvolume{110},
\bfpage{220502}
(\byear{2013})
\doiurl{10.1103/PhysRevLett.110.220502}
\end{barticle}
\endbibitem

\bibitem[\protect\citeauthoryear{Yang et~al.}{2018}]{Yang2018}
\begin{barticle}
\bauthor{\bsnm{Yang}, \binits{T.-S.}},
\bauthor{\bsnm{Zhou}, \binits{Z.-Q.}},
\bauthor{\bsnm{Hua}, \binits{Y.-L.}},
\bauthor{\bsnm{Liu}, \binits{X.}},
\bauthor{\bsnm{Li}, \binits{Z.-F.}},
\bauthor{\bsnm{Li}, \binits{P.-Y.}},
\bauthor{\bsnm{Ma}, \binits{Y.}},
\bauthor{\bsnm{Liu}, \binits{C.}},
\bauthor{\bsnm{Liang}, \binits{P.-J.}},
\bauthor{\bsnm{Li}, \binits{X.}},
\bauthor{\bsnm{Xiao}, \binits{Y.-X.}},
\bauthor{\bsnm{Hu}, \binits{J.}},
\bauthor{\bsnm{Li}, \binits{C.-F.}},
\bauthor{\bsnm{Guo}, \binits{G.-C.}}:
\batitle{{Multiplexed storage and real-time manipulation based on a multiple
  degree-of-freedom quantum memory}}.
\bjtitle{Nat. Comm.}
\bvolume{9}(\bissue{1}),
\bfpage{3407}
(\byear{2018})
\doiurl{10.1038/s41467-018-05669-5}
\end{barticle}
\endbibitem

\bibitem[\protect\citeauthoryear{Robertson et~al.}{2024}]{Robertson2024}
\begin{barticle}
\bauthor{\bsnm{Robertson}, \binits{E.}},
\bauthor{\bsnm{Esguerra}, \binits{L.}},
\bauthor{\bsnm{Me\ss{}ner}, \binits{L.}},
\bauthor{\bsnm{Gallego}, \binits{G.}},
\bauthor{\bsnm{Wolters}, \binits{J.}}:
\batitle{Machine-learning optimal control pulses in an optical quantum memory
  experiment}.
\bjtitle{Phys. Rev. Appl.}
\bvolume{22},
\bfpage{024026}
(\byear{2024})
\doiurl{10.1103/PhysRevApplied.22.024026}
\end{barticle}
\endbibitem

\end{thebibliography}
\newpage



\end{document}


\title{Supplementary: Distribution of light-matter quantum correlations with a temporally multiplexed solid-state quantum memory array}
\author[1]{\fnm{Aya} \sur{Mneimneh}}
\equalcont{These authors contributed equally to this work.}
\author[1]{\fnm{Susana} \sur{Plascencia}}
\equalcont{These authors contributed equally to this work.}
\author[1]{\fnm{Manuel} \sur{Gundin}}
\author[1]{\fnm{Jonathan} \sur{Hänni}}
\author[1]{\fnm{Samuele} \sur{Grandi}}
\author*[1]{\fnm{Markus} \sur{Teller}}\email{markus.teller@icfo.eu}
\author[1,2]{\fnm{Hugues} \sur{de Riedmatten}}
\affil[1]{\orgdiv{ICFO-Institut de Ciencies Fotoniques}, \orgname{The Barcelona Institute of Science and Technology}, \orgaddress{\city{Castelldefels (Barcelona)}, \postcode{08860}, \country{Spain}}}

\affil[2]{\orgdiv{ICREA}, \orgname{Institucio Catalana de Recerca i Estudis Avançats}, \orgaddress{\city{Barcelona}, \postcode{08015}, \country{Spain}}}

\date{\today}

\maketitle

\section{Autocorrelations and classical bound}

In Figs.~3 and 4, we compare the signal-idler correlations after storage in the QMA to the classical bound given by the Cauchy-Schwartz limit ${g_{s,i}^{(2)}=\sqrt{g_{s,s}^{(2)}g_{i,i}^{(2)}}}$.
We measure the signal and idler autocorrelations ($g_{s,s}^{(2)}$ and $g_{i,i}^{(2)}$) by inserting a \mbox{50-50} fiber beam splitter before the corresponding detectors and connecting the output ports to two signal detectors $s_{1,2}$ and two idler detectors $i_{1,2}$.
Following the experimental sequence outlined in Fig.~2, we conduct two sets of measurements: one in which telecom idler photons are detected at ICFO, and another in which they are detected at the Collserola Tower.
In the latter case, detection signals are recorded in the remote station, avoiding the need to send them back to ICFO.
We determine the coincidence probabilities $p_{x_1,x_2}$ of both detectors and the uncorrelated detection probabilities $p_{x_1}$ and $p_{x_2}$, with ${x\in\{s,i\}}$.

\begin{figure}
    \centering
    \includegraphics[width=1\linewidth]{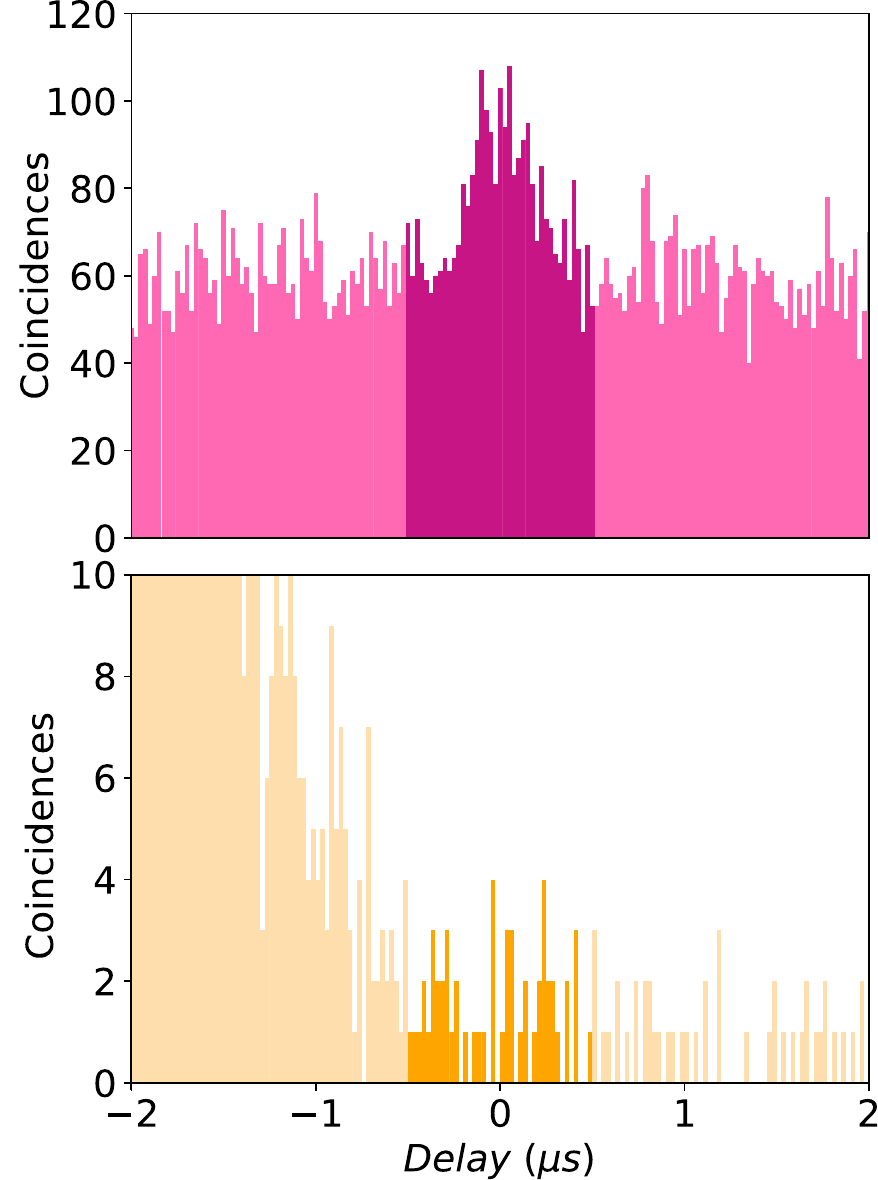}
    \caption{Coincidences of idler detections measured in Collserola (top) and signal detections measured at ICFO (bottom). The shaded areas indicate the detection window $T_\mathrm{DW}=\SI{1}{\micro\second}$.}
    \label{sup:autocorrelations}
\end{figure}

In Fig.~\ref{sup:autocorrelations} we plot the coincidence histograms for the two idler detectors~(top) at the Collserola Tower and for the two signal detectors (bottom) as a function of the delay between the two detection events. For the idler histogram, we find the coincidences peak around zero delay, as expected. 

Next, we evaluate the autocorrelations ${g_{x,x}^{(2)} = p_{x_1,x_2}/p_{x_1}p_{x_2}}$ within a detection window $T{_\mathrm{DW}=\SI{1}{\micro\second}}$, corresponding to the highlighted areas in Fig.~\ref{sup:autocorrelations}.
We obtain the idler autocorrelations $g_{i,i}^{(2)} = 1.24(2)$ for the local and $g_{i,i}^{(2)} = 1.27(3)$ for the distant measurements, and $g_{s,s}^{(2)}=1.7(3)$ for the signals.
These autocorrelations lead to the classical bounds $b_\mathrm{l}=1.4(1)$ for the correlations obtained locally and $b_\mathrm{d}=1.5(1)$ for correlations obtained at a distance.
Note that for Fig.~4, we repeat the outlined steps to evaluate $b_\mathrm{d}$ for the detection windows $T_\mathrm{DW}$ between 200 and $\SI{1200}{\nano\second}$. 

Finally, we evaluate the autocorrelation of two signal events detected in different storage trials $s$, following the procedure outlined in the main text for Fig.~3a. The resulting autocorrelations are plotted for trial shifts from $s=0$ to $5$ in Fig.~\ref{sup:autocorrelations}. While for the unshifted case $s=0$, we recover the previous value of $g_{s,s}^{(2)}=1.7(3)$, the autocorrelations decrease to values close to one for $s>0$, as expected from the fact that the signal events detected in different trials are uncorrelated. 


\begin{figure}
    \centering
    \includegraphics[width=1\linewidth]{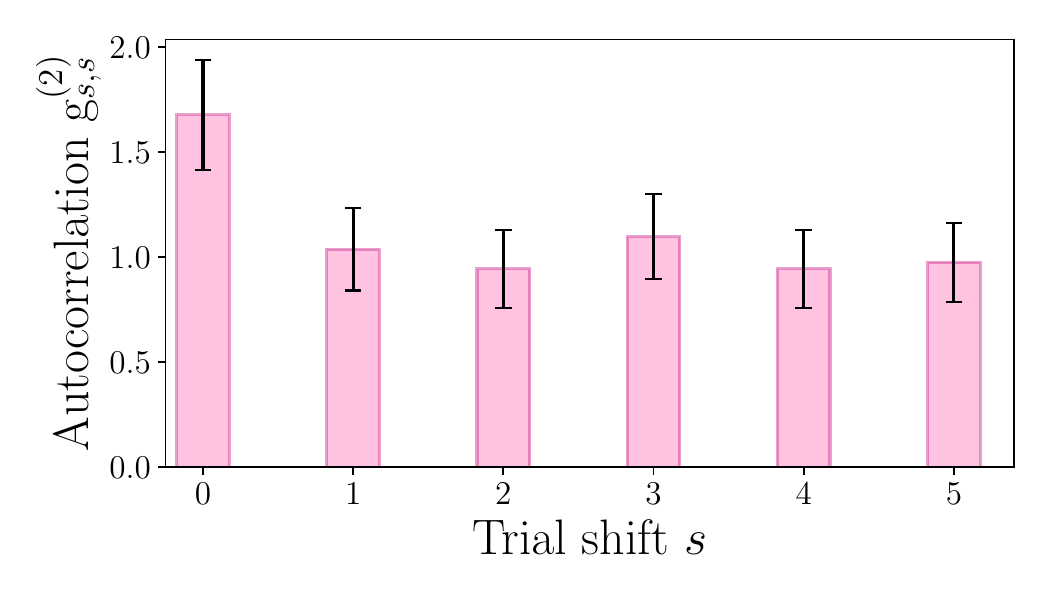}
    \caption{Autocorrelations of signals detected in a storage trial shifted by $s$.}
    \label{fig:comparison}
\end{figure}

\section{Limitations of quantum correlations}
In this section, we assess the limitations of the quantum correlations after storage in the QMA, discuss possible origins of these limitations and future mitigation strategies.

After quantum storage in the QMA, the cross-correlations of signal and idler are defined as~\cite{Albrecht2014}
\begin{equation}
    g_{s,i}^{(2)} = g_{AFC}^{(2)} \frac{SNR+1}{SNR+g_{AFC}^{(2)}} \label{eq:sup1}
\end{equation} with $SNR$ being the signal-to-noise ratio and $g_{AFC}^{(2)}$ describing the cross-correlation between idlers and signals stored only in the excited state, i.e., storage without on-demand control pulses. As a consequence of Eq.~\ref{eq:sup1}, the cross-correlations $g_{AFC}^{(2)}$  corresponds to an upper bound for $g_{s,i}^{(2)}$, as for $\lim\limits_{SNR \to \infty} g_{s,i}^{(2)} = g_{AFC}^{(2)}$.

To assess the limitations of the quantum correlations, we conduct two reference measurements performed in the laboratory at ICFO, from which we derive expected values of $g_{s,i}^{(2)}$ with Eq.~\ref{eq:sup1}.
To determine $g_{AFC}^{(2)}$, we employ the experimental sequence outlined in Fig.~2 without on-demand control pulses and follow the evaluation steps described in the main text to calculate the cross-correlation $g_{AFC}^{(2)}$ for each memory cell. The resulting values listed in Tab.~\ref{tab:ref_meas} range between $14(2)$ and $28(4)$ and significantly surpass the quantum correlations in Fig.~3b measured with on-demand control. 

To determine \textit{SNR}, we follow the approach of~\cite{Teller2025a} and replace the signal photons with weak-coherent inputs with an average photon number of $\bar{n}=0.61$ and a pulse shape that closely matches the shape of the signal photons. These inputs are immediately followed by the first CP of $\SI{4}{\micro\second}$ duration, such that the quantum memory output is emitted $\SI{6}{\micro\second}$ after the second CP. For each memory cell, we determine the signal counts $c_\mathrm{S}$ within a window $T_\mathrm{DW}$. We repeat these measurements without input pulses and determine the noise counts $c_\mathrm{N}$ in the same detection window. To account for the heralding efficiency $\eta_\mathrm{H}$ of the photon-pair source and for the multiplexing efficiency $\eta_\mathrm{M}$,  we rescale the signal counts to $\tilde{c}_\mathrm{S}=c_\mathrm{S}\frac{\eta_\mathrm{M}\eta_\mathrm{H}}{\bar{n}}$. The signal-to-noise ratio is then determined by
\begin{equation}
    SNR = \frac{\tilde{c}_\mathrm{S}-c_\mathrm{N}}{c_\mathrm{N}}
\end{equation}
The resulting values are listed in Tab.~\ref{tab:ref_meas} and range from $4(1)$ to $8(1)$.

\begin{table*}
    \centering
    \begin{tabular}{c|c|c|c|c|c|c|c|c|c|c}
       Cell  & 1 & 2 & 3 & 4 & 5 & 6 & 7 & 8 &9 & 10\\ \hline
             $g^{(2)}_{AFC}$  & 14(2) & 22(3) & 27(4) & 24(3) & 28(4) &17(2)  &17(2)  & 16(2) & 17(3) & 17(3)\\ \hline
       $SNR$ &4(1)   & 5(1) & 6(1) & 7(1) &8(1) & 7(1)  & 7(1) & 4(1) &7(1)  &4(1)\\ 

    \end{tabular}
    \caption{Reference measurements. Cross-correlations $g^{(2)}_{AFC}$ are obtained from experiments without CPs. The signal-to-noise ratio $SNR$ is determined from measurements with weak-coherent inputs.}
    \label{tab:ref_meas}
\end{table*}

With Eq.~\ref{eq:sup1} and the values of \textit{SNR} and $g^{(2)}_{AFC}$ we determine the expected values of the cross-correlation $g^{(2)}_{s,i}$. The resulting values are compared to the values obtained from the photon storage experiments of the main text in Fig.~\ref{fig:comparison}. For all memory cells, the expected values are higher than the values measured with single photons. 

We attribute this mismatch to higher noise in the photon storage experiments than in the reference measurements with weak-coherent inputs. We therefore compare the noise measured in both experiments in Fig.~\ref{fig:noise} per memory cell.
We find that the noise in the photon storage experiments from the main text indeed surpasses the noise from the reference measurement. 
We hypothesize that the increase in noise is due to two main reasons. First, the signal photons are detected within a $\SI{6}{\micro\second}$ window after the second CP, whereas the weak-coherent inputs are always detected $\SI{6}{\micro\second}$ after the second CP.
Fluorescence noise due to the CPs decays over time, and thus, signal photons detected earlier are subject to more noise. To quantify this effect, we evaluate the noise per herald in six steps of $\SI{1}{\micro\second}$ after the second CP for the photon storage measurements. In each step, we determine the noise within a window of $\mathrm{T}_\mathrm{DW}=\SI{1}{\micro\second}$ and afterwards determine the average over the six windows $\bar{c}_\mathrm{N}$. We find the noise to be on average $1.2$ times higher than the noise $c_\mathrm{N,6}$ measured in the detection window $\SI{6}{\micro\second}$ after the second CP. 

Second, the reference measurements were taken on different days with realignment of the optical system in between. Moreover, the reference measurements were performed with $\SI{4}{\micro\second}$-long control pulses, whereas the photon storage was performed with $\SI{3.5}{\micro\second}$-long control pulses for a longer photon acceptance window. To maintain the efficiency of the on-demand storage and retrieval, we increased the laser power of the control pulses with respect to the power in the reference measurements. We found in previous experiments that shorter CPs with higher power lead to an increase in noise. In conclusion, we attribute the fluorescence noise to be the limiting factor for the quantum correlations obtainable with our system.


\begin{figure}
    \centering
    \includegraphics[width=1\linewidth]{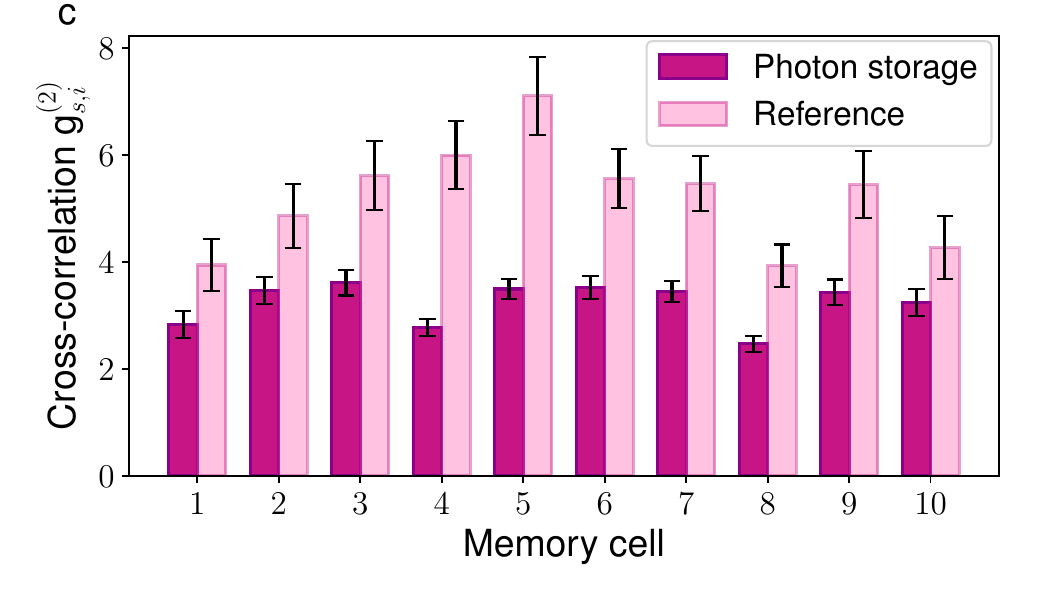}
    \caption{Comparison between the cross-correlations $g^{(2)}_{s,i}$ measured locally with spin-wave storage (labelled photon storage) and the expected values derived from the SNR and $g^{(2)}_{AFC}$ (reference).}
    \label{fig:comparison}
\end{figure}

\begin{figure}
    \centering
    \includegraphics[width=1\linewidth]{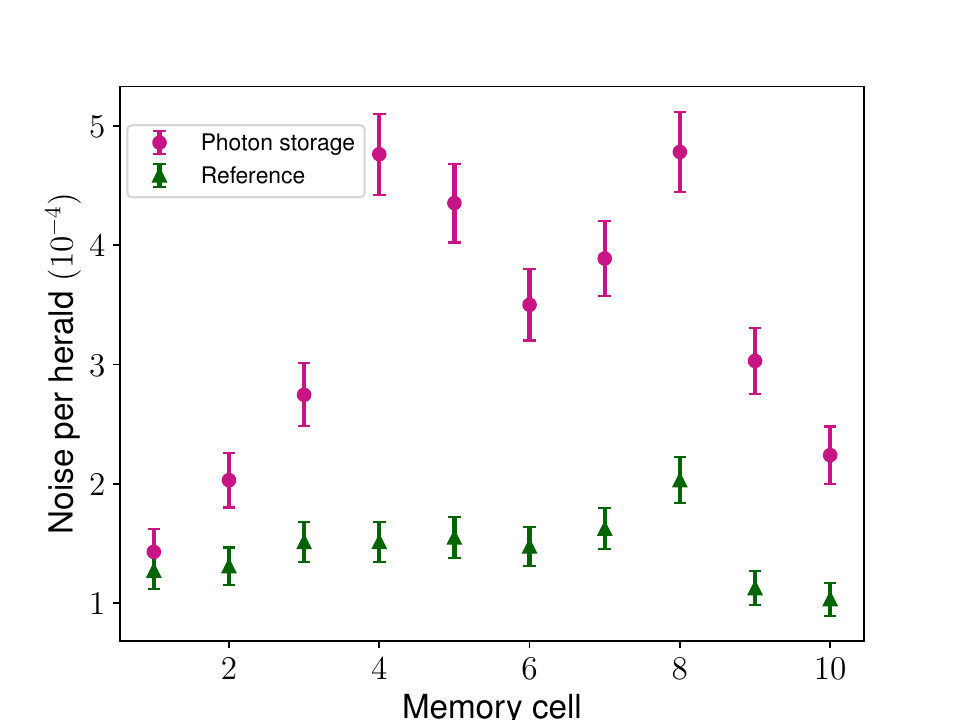}
    \caption{Noise per idler photon for each memory cell, compared between the spin-wave measurement (photon storage), a model taking into account the SNR (reference), and a model with additional noise (model).}
    \label{fig:noise}
\end{figure}

Currently, the quantum correlations obtained with the QMA are lower than those from comparable experiments with a single quantum memory cell in a crystal~\cite{Haenni2025}. We attribute the difference between these experiments to limitations in the quantum memory preparation: The distance between the quantum memory cells is $\SI{200}{\micro\meter}$ given by the frequency difference of $\SI{1}{\mega\hertz}$ between the radio-frequency tones sent to the AODs~\cite{Teller2025a}. This separation is comparable to the $4\sigma$ diameter of $\SI{200}{\micro\meter}$ of the preparation beam. Therefore, the preparation beam of each cell has non-zero overlap with the nearest neighbors. Due to the $\SI{1}{\mega\hertz}$ frequency difference, the preparation beams of the nearest neighbors affect the preparation process and lead to imperfect class cleaning. The separation between the cells was chosen as a trade-off between the AODs and fiber coupling efficiencies and the overlap of the preparation beams~\cite{Teller2025a}. With improved optical alignment of the AODs and fiber coupling, the separation between the cells could be increased to decrease the overlap while maintaining the outcoupling efficiencies.     

\section{Broadening of photons}
As visualized in Fig.4, selecting shorter detection windows increases the cross-correlations at the expense of coincidence rate, as only a fraction of the signal photon is considered. Narrowing the temporal size of the detected signal photons is therefore a path towards higher rates at high values of cross-correlations. In this section, we assess the current limitations and broadening mechanisms of the signal photon.

\begin{figure}
    \centering
    \includegraphics[width=1\linewidth]{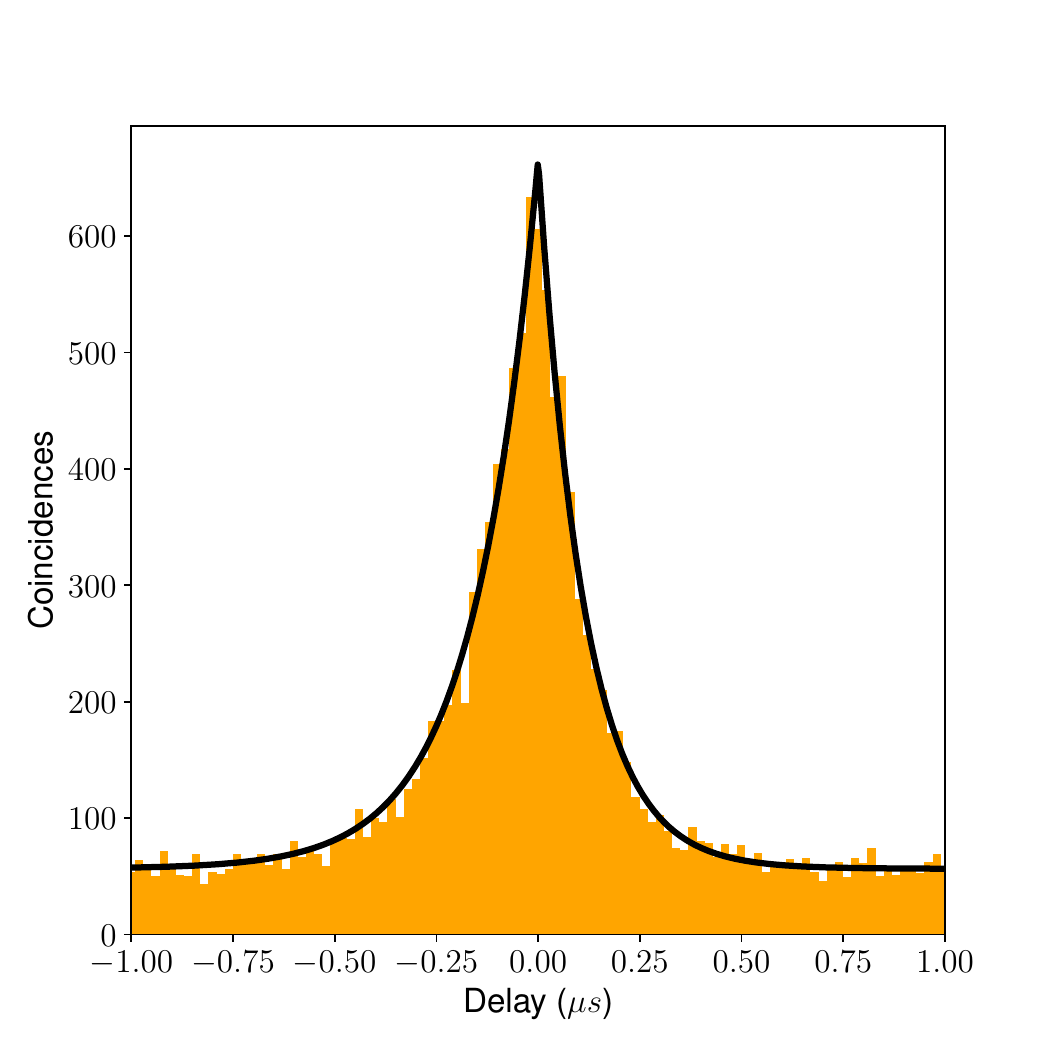}
    \caption{Coincidences of signal and idler detections as function of delay between detection events. Coincidences are measured bypassing the QMA and filtering stages. A FWHM of \SI{187(2)}{\nano\second} is determined from a fit of a double exponential.}
    \label{fig:sup_signal_idler}
\end{figure}

First, we measure signal-idler coincidences from photon pairs without passing through the QMA and the filtering stages. Both signal and idlers are detected at ICFO and coincidences between the two detectors are determined. In Fig.~\ref{fig:sup_signal_idler}, the coincidences are plotted as a function of the delay between the signal and idler detections. We fit the coincidence peak with a double exponential~\cite{Fekete2013} and extract idler photon linewidth of \SI{1.02(1)}{\mega\hertz} and a signal photon linewidth of \SI{1.40(2)}{\mega\hertz}, corresponding to a FWHM of \SI{187(2)}{\nano\second}. This value is determined by the resonator length of the photon-pair source and the reflectivities of the resonator mirrors~\cite{Fekete2013}.

\begin{figure}
    \centering
    \includegraphics[width=1\linewidth]{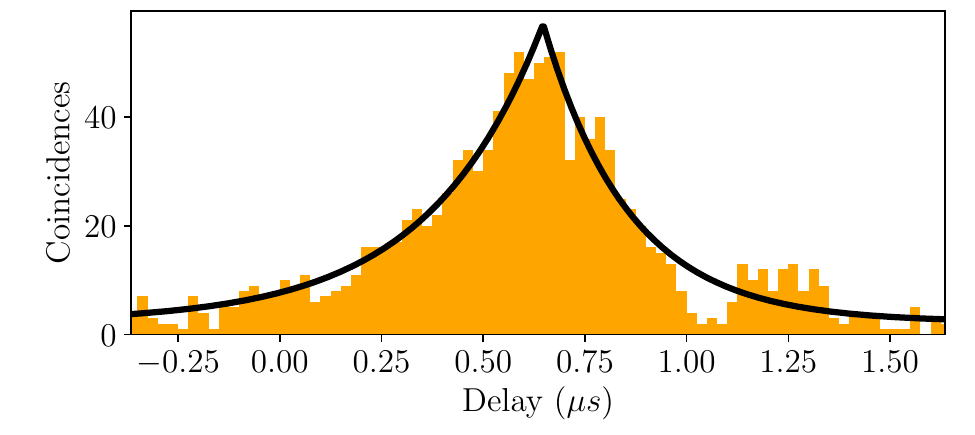}
    \caption{Coincidences of signal and idler detections as function of delay between detection events. Coincidences are measured after passing through the QMA and filtering stages. The QMA is prepared with a transparency window of $\SI{15.4}{\mega\hertz}$. The solid line corresponds to fit of to a double exponential with a FWHM of $\SI{341(12)}{\nano\second}$.}
    \label{fig:sup_pit}
\end{figure}

Next, we send the signal photons through the QMA, which is prepared with a transparency window of $\SI{15.4}{\mega\hertz}$, and the filtering stages. We repeat the previous steps and plot the coincidences of signal and idler detections in Fig.~\ref{fig:sup_pit}. We extract a FWHM of $\SI{341(12)}{\nano\second}$, which is significantly broadened from the value obtained without QMA and filtering stages.  We attribute this broadening to the filtering stage, in particular, to the filter crystal that is prepared with a transparency window of $\SI{2.4}{\mega\hertz}$. 
Moreover, a comparison to the FWHM of $\SI{531(13)}{\nano\second}$ obtained from the photon storage experiments (Fig.~3a) indicates that significant broadening results from the on-demand storage and retrieval process in the QMA.

In the future, the temporal width of the photons may be narrowed by adjusting the reflectivities and separation of the cavity mirrors of the photon-pair source. The broadening effects from the filter crystal may be reduced by employing the filter crystal in a double-pass configuration~\cite{Yang2018}, which may increase the filtering effect and allow consequently to widen the spectral filtering window. Machine-learning-based pulse shaping may be used to adjust the waveforms of the CPs for higher signal-to-noise ratios and lower photon broadening~\cite{Robertson2024}.   

\bibliography{qpsa.bib}
\newpage